\documentclass[twocolumn]{aastex701}

\usepackage{amsthm}
\usepackage{CJK}
\newtheorem{definition}{Definition}  
\usepackage{float}
\usepackage{natbib,aasdefs,url,bm}
\usepackage{graphicx}
\usepackage{color}
\usepackage{mathtools, epsfig, lipsum}
\usepackage{subfigure}
\usepackage{textcomp}
\usepackage{multirow,booktabs,colortbl,tabularx}
\usepackage{enumitem}
\usepackage{booktabs}
\usepackage{hyperref}
\usepackage{lineno}
\usepackage{silence}
\newcommand{\ESCf}{ESC_f}
\newcommand{\ESCs}{ESC_s}

\shorttitle{Beyond the Final Label: Improving Astronomical Light Curve Classification}
\shortauthors{Zhou, Malz, \& Schafer et al.}

\begin{document}
\begin{CJK*}{UTF8}{gbsn}
\title{Beyond the Final Label: Exploiting the Untapped Potential of Classification Histories in Astronomical Light Curve Analysis}

\author{Zhuoyang Zhou (周卓扬)}
\affiliation{Department of Statistics \& Data Science, Carnegie Mellon University, Pittsburgh, PA, USA}
\email[show]{grantz@andrew.cmu.edu}

\author{Alex I. Malz}
\affiliation{Space Telescope Science Institute, Baltimore, MD, USA}
\email{}

\author{Chad M. Schafer}
\affiliation{Department of Statistics \& Data Science, Carnegie Mellon University, Pittsburgh, PA, USA}
\affiliation{McWilliams Center for Cosmology and Astrophysics, Department of Physics, Carnegie Mellon University, Pittsburgh, PA, USA}
\email{}

\author{Konstantin Malanchev}
\affiliation{McWilliams Center for Cosmology and Astrophysics, Department of Physics, Carnegie Mellon University, Pittsburgh, PA, USA}
\email{}

\author{Guillermo Cabrera-Vives}
\affiliation{Department of Computer Science, Universidad de Concepci\'on, Concepci\'on, Chile}
\affiliation{Center for Data and Artificial Intelligence, Universidad de Concepci\'on, Concepci\'on, Chile}
\affiliation{Millenium Nucleus for Galaxies (MINGAL), Concepci\'on, Chile}
\email{}

\author{Christopher Hern\'andez}
\affiliation{Department of Physics and Astronomy, University of Pittsburgh, 3941 O’Hara St, Pittsburgh 15260, USA}
\email{}

\begin{abstract}
The Legacy Survey of Space and Time (LSST) on the Vera C. Rubin Observatory will generate a massive collection of time series (light curves) of the measured flux of transient and variable astronomical objects. 
With each new flux observation, light curve classifiers need to generate updated probability distributions over candidate classes, which will then be shared with the global community for the purpose of identifying interesting targets for follow-up observations as well as less time-sensitive analysis applications. 
Using the synthetic light curves and classification results of participating classifiers from the Extended LSST Astronomical Time-series Classification Challenge (ELAsTiCC), we investigate a novel framework to enhance existing light curve classifications by incorporating their classification histories and the temporal evolution of these histories. 
To demonstrate the potential of this approach, we introduce a model that combines a recurrent neural network and an additive attention module, which shows improved classification accuracy and more balanced precision-recall performance compared to existing classifiers from the challenge. 
Furthermore, at this stage, most, if not all, of the existing classifiers are evaluated by their final classification results on complete light curves;
we propose new metrics that evaluate the stability, accuracy, and early classification performance of a classifier's predictions when using limited data by considering the Wasserstein distance between the temporally evolving classification probability distributions. 
Our metrics offer a more comprehensive perspective for model assessment by supplementing classical methods such as the confusion matrix and precision-recall. 
\end{abstract}


\section{Introduction}\label{sec:intro}

The next-generation sky survey Legacy Survey of Space and Time (LSST; \citealp{abell2009lsst}), to be conducted on the Vera C. Rubin Observatory, will repeatedly scan the sky for 10 years and provide temporal astrometric and photometric data for 20 billion objects. 
The repeated scanning enables the telescope to detect objects with changing flux by comparing the captured images with a reference template. 
The result will be time series (light curves) of flux measurements observed through six broadband photometric filters (passbands) for 1-2 orders of magnitude more transients and variable stars than have been observed to date \citep{ivezic2019lsst, hlovzek2023results}. 
The LSST Data Management (DM) system will produce two related data products: 
(1) real-time alerts of flux observations for detected objects by continuously processing the incoming stream of images captured by the observatory, and 
(2) annual static Data Releases (DRs) containing unlabeled light curves from all survey data since initiation \citep{abell2009lsst}. 
The survey itself will mainly process and distribute the acquired flux observations and light curves, leaving the production of real-time classifications of these data products to community alert brokers.

To catalyze the development of classifiers of transient and variable objects, public and community-wide data challenges such as the Photometric LSST Astronomical Time-Series Classification Challenge (PLAsTiCC; \citealt{hlovzek2023results}) and the Extended LSST Astronomical Time-series Classification Challenge (ELAsTiCC) were conducted using synthetic data that mimics the observation scheme of the real LSST survey. 
PLAsTiCC was launched as a public Kaggle challenge\footnote{\url{https://www.kaggle.com/c/PLAsTiCC-2018}} designed to mimic the real LSST light curve observations with a test set of 3.5 million light curves with over 453 million photometric measurements, and a training set containing 8,000 objects. The significantly smaller training set was intended to simulate the challenges of incompleteness and non-representativity that arises in classifying LSST light curves. This was developed by members of the Transient and Variable Stars Science Collaboration (TVS) and the Dark Energy Science Collaboration (DESC).
PLAsTiCC represents the mainstream paradigm for developing light curve classifiers, where the model is trained and evaluated on static and complete light curves. 
With over 1,000 teams participating, the challenge produced classifiers employing a diverse set of techniques, ranging from boosted decision trees to neural networks, notably yielding classification probability mass functions (PMFs) rather than deterministic class predictions. 
The top three classifiers made significant improvements over the state-of-the-art within the astronomy community in the classification of Type Ia supernovae and kilonovae \citep{hlovzek2023results}.

While PLAsTiCC achieved great success as one of the most complete simulations of future photometric surveys, its design failed to account for the real-time alert classification scheme. 
In real operation, new detections are streamed from the LSST Data Management System to the real-time community alert brokers, whose classifiers will make predictions with ongoing, partial light curves that are updated nightly. 
In practice, each new observation of a detected object will result in an updated PMF over candidate classes. These PMFs will then be globally disseminated for the purpose of identifying interesting targets for detailed follow-up observations and myriad downstream analyses. 

The Extended PLAsTiCC (ELAsTiCC) was designed to stress-test such end-to-end real-time pipelines for light curve classification. 
Instead of directly providing complete light curves, ELAsTiCC streamed detections as alerts through the cloud-hosted ZTF Alert Distribution System (ZADS; \citealt{patterson2018zwicky}). 
Classifiers were required to ingest these real-time alerts through various delivery mechanisms, such as Apache Kafka and Google Pub/Sub, and produce classifications for every new detection rather than a single final classification on the complete light curve per object. 
ELAsTiCC was primarily targeted at alert brokers, which are professional automated software systems and classifiers developed through research collaborations to process, characterize, and prioritize alerts from the LSST and other surveys for follow-up observations. 
Participating alert brokers included ALeRCE \citep{forster2021automatic}, FINK \citep{2021MNRAS.501.3272M}, Pitt-Google \citep{pittGoogle}, ANTARES \citep{matheson2021antares}, and AMPEL \citep{nordin2025ampelworkflowslsstmodular}.

Classifiers applied to both the PLAsTiCC and ELAsTiCC challenges, as well as light curve classifiers in general, typically either directly use the full light curves with deep learning models \citep{boone2021parsnip, pasquet2019pelican, cabrera2024atat} or extract informative features for training other machine learning models like tree-based classifiers such as random forests or boosted decision trees \citep{boone2019avocado, de2024superphot+, sanchez2021alert, matheson2021antares}. A wide range of approaches has been implemented in automated pipelines for light curve classification, leading to major improvements in accuracy and rare event identification \citep{pruzhinskaya2026anomaly}.
On the other hand, efforts to tackle the light curve classification in an incremental learning style with data arriving sequentially have been less prevalent due to the increased challenge of the work and computational constraints on experimental design for validation. 

For real-time alerts, the approach taken by the majority of the state-of-the-art light curve classifiers can be characterized as follows: 
Whenever a new observation of a detected variable or transient object is made, the classifiers refit the enlarged full light curve to generate updated probability distributions over candidate classes.
Such a scheme suffers from the limitation that it ignores the entire classification history that was built from earlier alerts with partial light curves.
By constructing a meta-classifier that ingests both the raw light curves and the classification histories from the base model, these classification histories, as time series of classification PMFs, can improve the classification in several ways: 

\begin{enumerate}
    \item \textbf{Stacking with enriched feature space}: 
    Our approach can be viewed as an application of {\it stacking} \citep{wolpert1992stacked} classifiers sequentially rather than in parallel, an approach also known as the cascade generalization \citep{gama2000cascade}. 
    The classification history provides an expanded feature space where the base classifier acts as a feature extractor, capturing patterns in a representational language potentially different from that of the meta-classifier. 
    This architectural diversity enables the two models to complement each other, yielding improved performance (e.g., decision trees for the base model and neural networks for the meta-model).
 
    \item \textbf{Self-reflective error/bias correction}: 
    The classification histories can be considered to be soft labels with uncertainty-annotated suggestions. 
    The meta-classifier can recalibrate the base classifications by rectifying systematic classification bias and errors by learning mistake patterns of the base model and its overconfidence/underconfidence behavior throughout the historical classifications. 
    
    \item \textbf{Following an incremental learning style}: 
    By producing new classifications conditioned on previous classification histories, the meta-classifier naturally resembles an incremental learning style. 
    Taking into account both recent and earlier classifications could achieve robustness against short-term noise and flux fluctuations, whether caused by intrinsic variability of astronomical objects or observational uncertainties, resulting in a more stable classification.
\end{enumerate}

This work also seeks to develop and test related performance metrics. 
Though metrics of PMFs have been developed \citep{malz2019photometric} and applied \citep{hlovzek2023results} to PLAsTiCC, existing quantitative approaches are evaluated based on their final classification probabilities obtained with full light curves using classic model assessment tools, including the classification accuracy, confusion matrix, and other measures like the Receiver Operating Characteristic (ROC) curves and F1 scores. The model development and training processes are also mainly tailored for optimizing these metrics. 
These evaluation strategies are insufficient for assessing the probabilistic classification of real-time alerts and informing trustworthy decision-making. 

In particular, current assessment methods do not fully evaluate both classification stability and early classification performance. First, an ideal classifier should not only give the correct classification, with high classification probability assigned to the true class, but also maintain stable and consistent classifications over time rather than frequently switching between favored classes as new observations arrive. Since follow-up observations with spectrographs are resource-intensive, unstable classifiers can be less trustworthy when used in decision-making; the classification stability\footnote{Here stability always refers to the assigned classification rather than inherent instability of physical class, which may occur for some variable sources that transition between states.}, which is not reflected by classical metrics, can be effectively measured with classification histories from partial light curves by imposing a valid distance measure between classification PMFs. 

Second, another important classifier characteristic that is often neglected by classical metrics is the early-classification performance. 
Transient objects, like kilonovae, can have very short life spans, and quick follow-up analysis can be extremely valuable. 
This motivates an evaluation metric that accounts for the time or number of observations the classifier takes to obtain a correct and stable classification. 
Together with the stability requirement, an ideal classifier should use fewer observations or time to achieve stable correct classification with a high classification probability assigned to the true class. 

These limitations highlight the importance of classification histories for both enhancing the classification performance and giving a more comprehensive model evaluation that better informs decision-making in analyzing real-time alerts. 

This work aims to address these limitations efficiently by leveraging the classification histories in both classifier training and evaluation. 
We use the synthetic data and classification results of participating classifiers from ELAsTiCC\footnote{The ELAsTiCC data is available at \url{https://portal.nersc.gov/cfs/lsst/DESC_TD_PUBLIC/ELASTICC/}} to investigate a new framework for enhancing existing light curve classifiers by supplementing the raw flux observations with their classification histories. 
In particular, we propose a model that incorporates a Long Short-Term Memory (LSTM) \citep{hochreiter1997long} network for processing the combined multivariate time series of raw flux and classification histories with associated temporal features derived from raw timestamps. 
We apply an additive attention mechanism \citep{bahdanau2014neural} to combine the sequence of hidden states from the LSTM by selectively weighting the importance of different timestamps. 
We evaluate the new framework by selecting three representative baseline classifiers from ELAsTiCC with reasonably good classification coverage, focusing on the five common supernova classes, namely type Ia, Ib/c, II, Iax, and 91bg. 
The new models show higher classification accuracy when compared with each of the selected classifiers and show more balanced precision-recall performance.

We also propose new model performance metrics and visualizations, named Early-Stable Classification Metrics, that can better assess the classifier's stability and early classification ability by systematically evaluating its classification histories. 
The metrics quantify the temporal evolution of classification PMFs and use the Wasserstein distances \citep{kolouri2017optimal, peyre2019computational} between classification PMFs for stability assessment. 
We propose a convergence-time version of the new metric that offers a more intuitive understanding of classifier performance and a weighted-sum version that gives a concise summary and can be potentially adapted to a loss function for classifier training and optimization. 
We demonstrate some potential applications of the new metrics by applying them to both our proposed and existing classifiers from ELAsTiCC. 
Our proposed models also demonstrate superior early-stable classification performance when compared with the selected base models. 

The paper layout is as follows: 
We introduce the background related to LSST and existing progress and limitations of light curve classification in Section~\ref{sec:backgrounds}. 
The ELAsTiCC data and the data engineering procedure are introduced in Section~\ref{sec:data}. 
In Section~\ref{sec:methods}, we introduce the proposed classifier architecture and new metrics for model assessment. 
In Section~\ref{sec:experiments}, we present the model training process and experimental results that compare the proposed classifier with selected base classifiers from ELAsTiCC using both classical and new metrics. We also demonstrate the application of the new metric for model assessment with visualizations. 
Discussion and future directions are given in Section~\ref{sec:discussion}.

\section{Background}\label{sec:backgrounds}

One of the primary scientific objectives of the LSST is to greatly extend our knowledge of transient and variable stars using photometric observations. 
By repeatedly scanning the night sky and taking observations in the $u, g, r, i, z$, and $y$ passbands, the Vera C. Rubin Observatory will generate a large collection of light curves of measured flux for detected objects by comparing the captured images with a reference template that was built from previous observations \citep{abell2009lsst}. 
Compared with previous large-scale surveys such as the Sloan Digital Sky Survey (SDSS; \citealp{york2000sloan}) and the Two-Micron All Sky Survey (2MASS; \citealp{2006AJ....131.1163S}), LSST will cover a much larger survey area and depth, generating around 30 terabytes of processed data each night and a trillion-line database with temporal astrometric and photometric data on 20 billion objects over a survey period of ten years \citep{abell2009lsst}. 

\subsection{Data Products and Synthetic Light Curves}

The LSST DM System will produce two main data products.
First, the system will continuously process the incoming stream of images captured by the observatory and produce real-time alerts of flux observation, which leads to a constantly updated light curve for detected objects. 
These real-time alerts are crucial for informing the choices of quick follow-up observations of transient objects with very short life spans, rarely observed events, and discovering unanticipated phenomena. 
Second, roughly once per year, the observatory will produce a static DR that consists of unlabeled light curves for detected objects generated from all survey data taken from the date of survey initiation to the cutoff date for the DR \citep{abell2009lsst}. 
The DR is more suitable for less time-sensitive studies, such as building catalogs with labeled data and population-level analyses. 
Both data products require accurate and reliable classification, with an extra emphasis on classification stability and early classification performance for the real-time alerts. 

\begin{figure*}
\centering
\includegraphics[width= 0.85 \linewidth, scale=0.4] 
{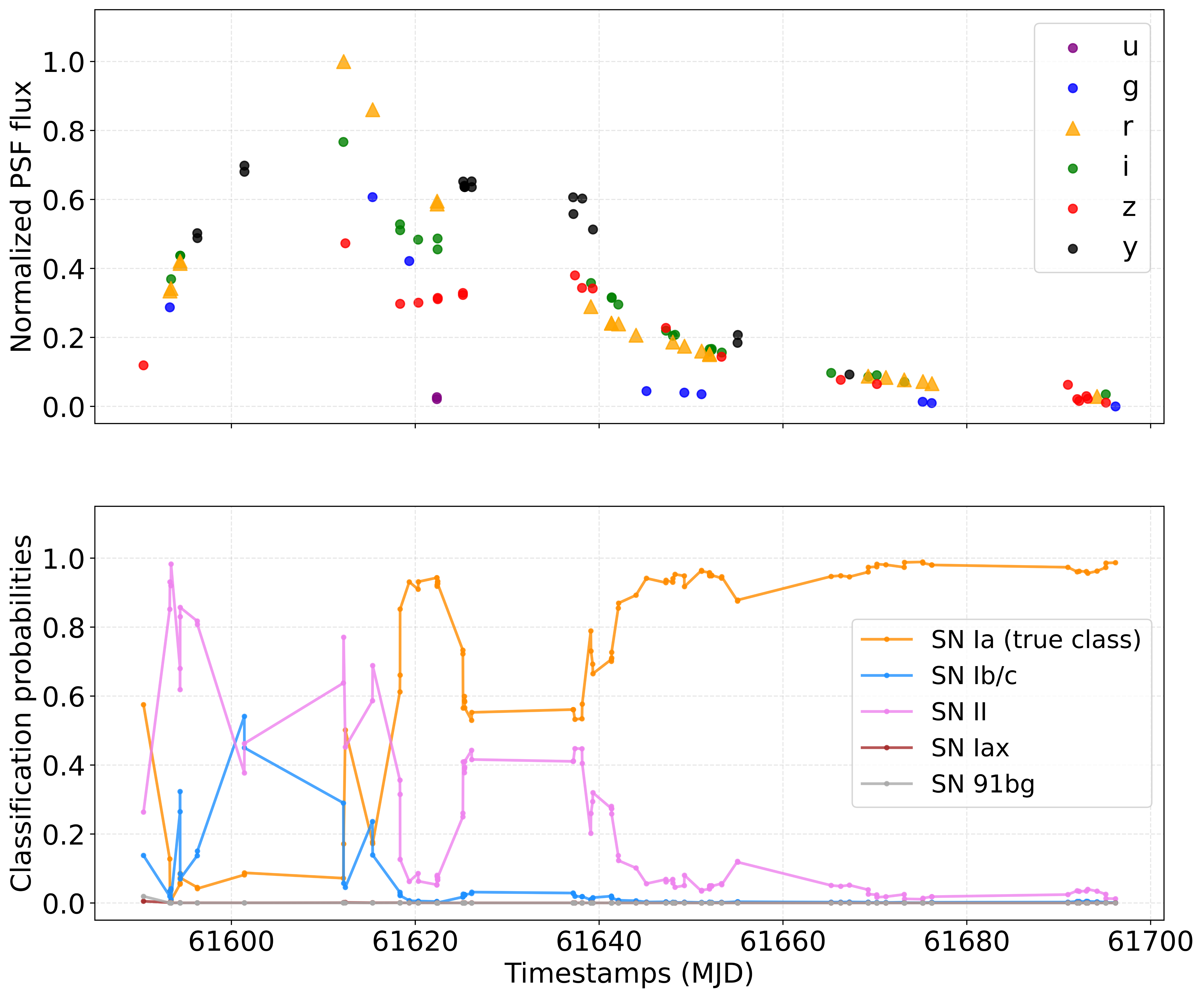} 
\caption{\label{fig:lc_example} Synthetic light curves in six passbands (top) and classification results (bottom) of a Supernova Type Ia (SN Ia) with object\_id=100971671 from ELAsTiCC2. The classification results are from one of the participated classifier with name hidden by anonymity requirements. The \textit{r}-band flux (orange triangles) exhibits a clear peak followed by a decline. The classifier achieves a high confidence in the true class upon observing the rising flux, peak, and the subsequent decay to low luminosity after MJD 61640. Observational uncertainties are omitted for clarity.}
\end{figure*}

For illustration, in Figure \ref{fig:lc_example} we plot the complete synthetic light curves of a Supernova Type Ia (SN Ia) (object\_id=100971671) observed in six passbands from the ELAsTiCC2 \citep{hlovzek2023results} simulation. We also provided the full classification results for each flux observation from one of the participated classifier with name hidden by anonymity requirements.
The primary characteristic of a Type Ia supernova is that it initially remains at a lower luminosity and then undergoes an explosion that results in a drastic increase in flux, followed by the decay to low luminosity again. 
This trend is reflected in provided light curve and especially clear in the \textit{r}-band light curves (orange triangles) in Figure \ref{fig:lc_example}, where there is a flux decay from a significantly higher flux between 61600 and 61620 MJD. 
Note that due to statistical fluctuations and the method used to estimate brightness, the flux may be negative for dim sources, particularly when the true flux is close to zero \citep{allam2018photometric}. 

LSST light curves present features that complicate the classification objectives. 
Since the telescope does not continuously observe the same region of the sky, observations may contain gaps ranging from minutes to days for short interruptions and several months for seasonal visibility changes. 
Figure \ref{fig:lc_example} illustrates this with irregular gaps between the clusters of flux observations. This also suggest that objects are only partially observed. For example, while we have a relatively complete observation here, we may only detect the decaying phase of luminosity while missing the initial rise stage for some other objects.
Further complicating the problem, while there are six passbands, only two passbands will be used for sequentially imaging the observed area. One of these two passbands will then record a follow-up observation of the same area several hours later. This creates a multivariate time series classification task where different channels have missing data at different time points. 
Combined with instrumental and observational uncertainties, these factors make LSST light curves highly irregular multivariate time series with heteroscedastic errors and profound missingness patterns that lack any regularity.

\subsection{Existing Methods and Classification Scheme}

Existing frameworks and classifiers for light curve classification mainly use machine learning-based methods instead of classic parametric time series models.

There are three main types of modeling approaches: 
(1) directly using the raw light curves as inputs for deep learning models, 
(2) pre-processing the light curves and extracting informative features for training other supervised machine learning models, like a tree-based classifier, such as random forests or boosted decision trees, or 
(3) combining both in a more complicated classification pipeline. 
We provide a brief and non-exhaustive overview of some representative classifiers.

Representative classifiers built on deep learning include the Pelican (deeP architecturE for the LIght Curve ANalysis; \citealt{pasquet2019pelican}) and ParSNIP (Parameterization of SuperNova Intrinsic Properties; \citealt{boone2021parsnip}). 
Pelican transforms light curves into 2D ``light curve images'' (LCIs), where the width represents the temporal axis and the height represents the photometric passbands, then uses a convolutional neural network (CNN) for classification. 
ParSNIP trains a generative model directly from large samples of light curves that combines a modified version of a variational autoencoder (VAE; \citealt{kingma2013auto}) and an explicit physics-based model that informs how light from the transients propagates and is observed. 
The representation learned by the ParSNIP model can be used for light curve reconstruction or feature inputs for light curve classification. 

For the second approach, some representative classifiers include the winner of the PLAsTiCC \citep{hlovzek2023results}, Avocado \citep{boone2019avocado}, which used a Gaussian process regression for light curve interpolation followed by feature extraction for tree-based models such as gradient-boosted decision tree and random forest. 
Other classifiers include Superphot+ \citep{de2024superphot+} and ALeRCE \citep{forster2021automatic}, which also used feature extraction followed by neural network or random forest-based classifiers. 

AMPEL \citep{nordin2025ampelworkflowslsstmodular} is an example of an approach that uses a more complicated framework that combines multiple classifiers into a single pipeline. 
This framework provides different models that are specifically designed for rapid identification with ongoing observations or classification of more complete light curves. 
The modeling approach consists of both feature extraction, followed by tree-based models, or using light curves as direct input by leveraging the ParSNIP model mentioned above. 

\section{Data and Analysis Set-up}\label{sec:data}

In this section, we briefly introduce the main synthetic datasets that were crafted specifically for the advancement of LSST light curve classification methods, ELAsTiCC \citep{elasticc}. We also introduce the data engineering procedures applied to the ELAsTiCC data.

\subsection{ELAsTiCC}

To increase the realism for classifiers and to test the full data processing pipelines mimicking the real LSST classification workflow, the LSST DESC later created the Extended LSST Astronomical Time-series Classification Challenge (ELAsTiCC; \citealt{elasticc}). 
Instead of being launched as a public data challenge, ELAsTiCC was designed primarily for alert brokers to test their end-to-end real-time pipeline for light curve classification. 
While there are two data challenges, ELAsTiCC and ELAsTiCC2, we mainly focus on the ELAsTiCC2 data, which uses a more current simulated LSST cadence that better reflects the anticipated LSST observations and has better coverage in terms of classifications from broker teams.

ELAsTiCC2 simulated light curves for 4.1 million transient and variable stars. These light curves are sampled using a simulated LSST baseline 3.2 cadence (observation strategy) in order to mimic the real observation scheme. 
This yields 62 million detections and 990 million photometry measurements. 
Unlike the PLAsTiCC, which directly provided the full three-year light curves, ELAsTiCC streamed these detections as alerts to the cloud-hosted ZTF Alert Distribution System (ZADS; \citealt{patterson2018zwicky}). 
Five alert brokers participated in the ELAsTiCC2, including ALeRCE \citep{forster2021automatic}, FINK \citep{2021MNRAS.501.3272M}, Pitt-Google \citep{pittGoogle}, ANTARES \citep{matheson2021antares}, and AMPEL \citep{nordin2025ampelworkflowslsstmodular}. 
Alert brokers ingested these real-time alerts through various delivery mechanisms, including Apache Kafka and Google Pub/Sub. 
The brokers then applied classifiers that were trained on a separate dataset provided by the challenge. 
The result was object classifications for each new alert, which were then submitted to DESC for evaluation. 
More detailed documentation regarding the data generation process, classification taxonomy, original training data, and truth table can be found in \cite{elasticc}.

\begin{figure*}
\centering
\includegraphics[width= 0.7 \linewidth, scale=0.15]{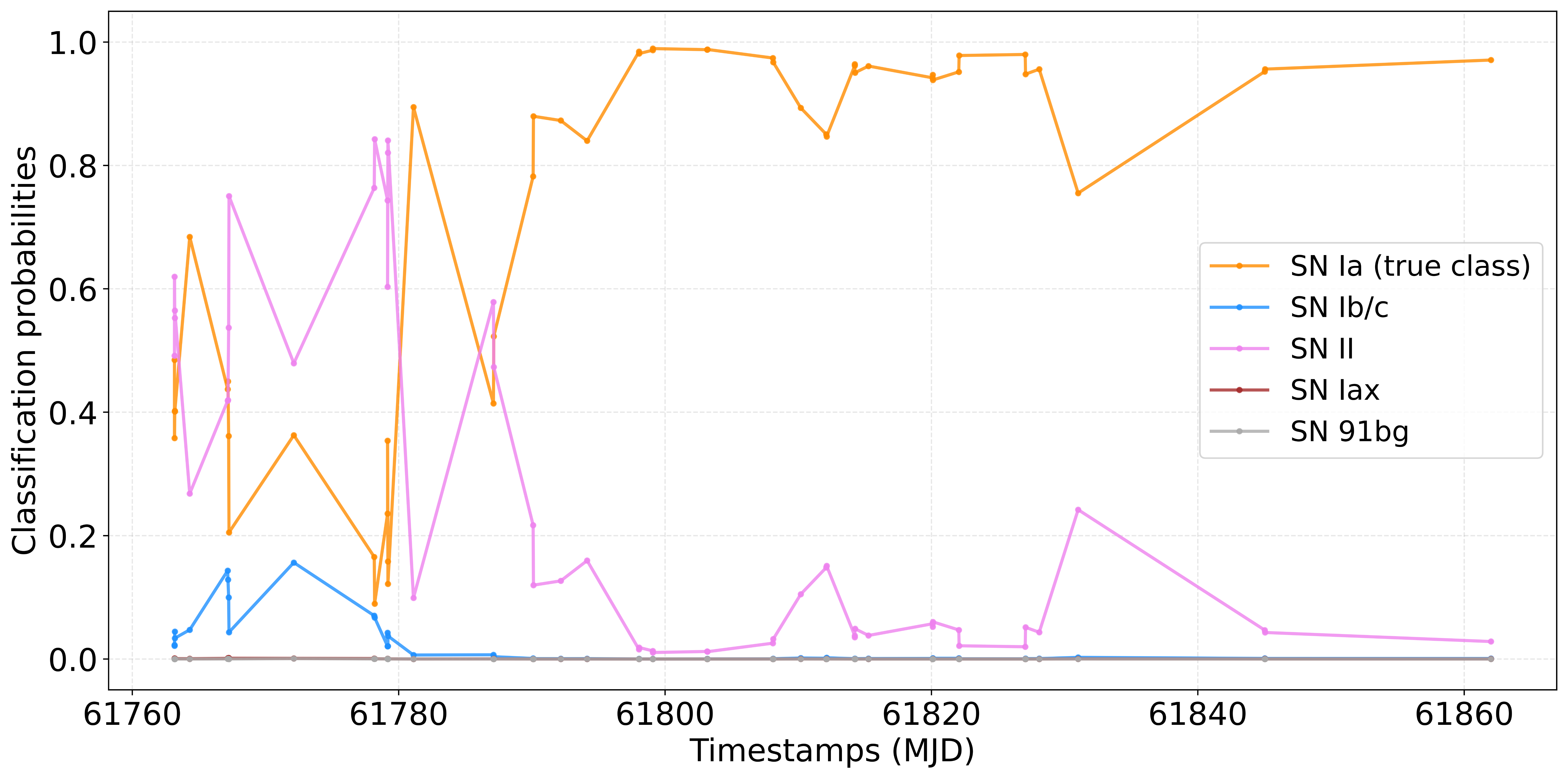}  
\includegraphics[width= 0.7 \linewidth, scale=0.15]{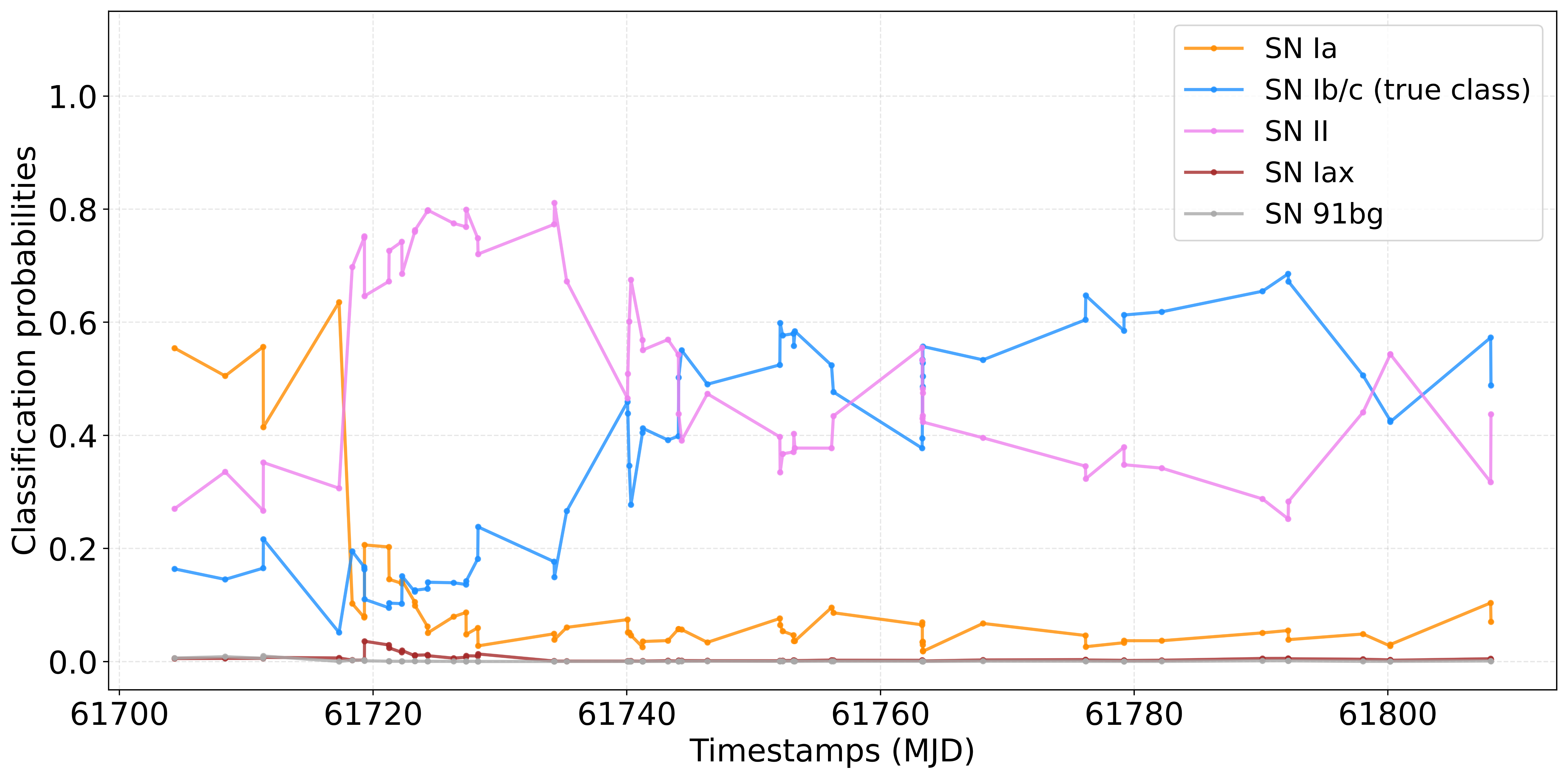}  
\caption{\label{fig:pmfs} 
Time series of classification PMFs from classifier A for a SN Ia (upper, object\_id=10362584) and SN Ib/c object (bottom, object\_id=1472297). 
For the SN Ia object, the classifier A made a relatively satisfactory classification with high and stable (despite one downward jump) classification probability for the true class after MJD 61800. 
For the SN Ib/c object, the classifier exhibits confusion between SN II and SN Ib/c classes, characterized by more substantial variations in classification probabilities, ultimately resulting in a correct but ambiguous classification.
}
\end{figure*}

\subsection{Analysis Set-up}

Instead of developing new classifiers from scratch, the main focus of this work is to investigate how to enhance existing classifiers and to provide more informative model assessment with their classification histories. 
To this end, we obtained complete classification results, in the form of time series of classification PMFs, from all participating ELAsTiCC2 classifiers. 
While the ELAsTiCC2 provides a hierarchical classification taxonomy, with 25 classes in total, we focus the demonstration of our methodology on the five most common supernova classes, including SN Ia, SN Ib/c, SN II, SN Iax, and SN 91bg. 
(This selection creates a class imbalance with SN Iax and SN 91bg as minority classes.) 

Our choice is mainly motivated by several considerations. 
First, we want to avoid a mix of variable and transient objects, which have fundamentally different temporal behaviors. 
Second, many rare event types such as kilonovae and dwarf novae contain too few objects with sufficient classification histories to enable the learning of temporal patterns. 
For instance, most kilonovae have fewer than five classifications per object. 
Finally, supernova classification itself represents a particularly demanding challenge. Even among the most common types, subtle photometric similarities can make discrimination difficult, especially at early stages when few classifications are available. The real-world detections are noisy, incomplete, irregular as shown in Figure \ref{fig:lc_example} and required timely and accurate classification given the transient nature of these sources.

These complexities therefore make classification among the five most common supernova types a meaningful benchmark for assessing the capacity of our approach: it provides a sufficient starting point for evaluating our framework and its potential to extend more broadly for other object types while maintaining sufficient sample sizes across all classes. 

Among the alert brokers that participated in the challenge with available classification results, we chose three representative classifiers with relatively high classification coverage and good classification performance. 
Due to anonymity requirements, we cannot provide identification information for the classifiers and simply refer to them as classifiers A, B, and C, where classifiers A and B are from the same alert broker, and classifier C is from another broker. 
Some of these classifiers are already adopted in mature classification pipelines that are publicly available. 
The detailed classification performance of the selected classifiers is evaluated in Section \ref{sec:experiments}. 

Since classifiers only labeled a subset of all available test objects, we only include objects for which a classification is available for each classifier. Classifier C has different coverage than that of classifiers A and B. More details of the coverages can be found in Table \ref{table1}. 
Classifiers A and B labeled approximately 25\% of all ELAsTiCC2 synthetic objects, while Classifier C achieved substantially higher coverage at approximately 70\%. 
We exclude the remaining classifiers for three main reasons: 
they are duplicates from the same alert broker (revised partway through the challenge for technical reasons), they have relatively poor coverage (classifying only a few percent of objects), or they fail to produce meaningful classifications (e.g., maintaining near-zero or constant probabilities across most objects).

For each object, we have a time series of synthetic flux, classification PMFs over the five classes, and observation timestamps. 
Other features, including the corresponding passbands of the flux and flux-to-error ratio, will not be used for modeling. 
In Figure \ref{fig:pmfs}, for illustration, we plot the classification PMFs from classifier A for one SN Ia object and one SN Ib/c for the five common supernova classes. 
For the SN Ia object, the classifier ultimately achieved a satisfactory performance. 
After the initial unstable stage, it reached a high classification probability for the true class after MJD 61800. 
The classifications are also fairly consistent without large variations. 
This would be considered a relatively ideal classification that has good early classification performance and classification stability. 
On the other hand, for the other SN Ib/c object, the classifier demonstrates relatively poor performance with a correct but very ambiguous final classification and substantial confusion between SN Ib/c and SN II classes. 
One class repeatedly overtook and was overtaken by another after MJD 61740, indicating fairly unstable classifications. 
Despite the correct final classification, the probabilities for both the SN Ib/c and SN II classes are near 0.5, making the result less informative for decision-making at any time before stability is achieved.

\begin{table*}
\centering
  \caption{Summary of Sample Sizes after Filtering and Classification Coverages}
  \label{table1}
  \setlength{\tabcolsep}{4pt}
  \begin{tabular}{l|cc|cc|cc}
    \toprule
    & \multicolumn{2}{c|}{Classifier A \& B} & \multicolumn{2}{c|}{Classifier C} & \multicolumn{2}{c}{ELAsTiCC2} \\
    \cmidrule(lr){2-3} \cmidrule(lr){4-5} \cmidrule(lr){6-7}
    Classes & $\geq15$ (\% retained) & Full (coverage) & $\geq15$ (\% retained) & Full (coverage) & Total  & Median Length \\
    \midrule
    SN Ia& 93771 (21.54\%) & 435277 (26.9\%) & 121869 (10.56\%)& 1153596 (71.28\%) & 1618391 & 4\\
    SN Ib/c & 21433 (24.73\%) & 86670 (27.46\%) & 30328 (13.51\%) & 224462 (71.13\%) & 315549 & 4 \\
    SN II & 79290 (18.75\%) & 422941 (28.3\%)  & 114764 (11.1\%) & 1034253 (69.2\%) & 1494484 & 3\\
    SN Iax & 2039 (18.15\%) & 11232 (26.34\%) & 2679 (8.82\%) & 30363 (71.2\%) & 42644 & 3\\
    SN 91bg & 1545 (18.73\%) & 8247 (23.83\%) & 1949 (7.67\%) & 25415 (73.44\%) & 34606 & 4\\
    \bottomrule
  \end{tabular}
\end{table*}

Since the number of classifications per object ranges from a few to more than a thousand, to have enough sample size for studying the temporal evolutions and to prevent objects in the same training batch from differing drastically in length, we only select objects with at least 15 classifications and apply truncation with a maximum length of 100 timestamps. 
Note that one could use different lower and upper cuts for a more holistic investigation. 
The sample sizes after the filtering, labeled as percentages retained, are presented in Table \ref{table1} for all selected classifiers. 
The median length of classifications for all classes is less than five, which explains the relatively low retention percentages.

For most objects, observations are irregularly spaced and clustered, including some with large gaps with no measurements, due to the survey's observation plan. The irregular nature is well-reflected in Figure \ref{fig:pmfs}. 
To account for the highly irregular time series, we compute four additional absolute and relative temporal features based on the observation history for each object, namely, the raw timestamps, time since the first classification, time since the previous classification, and time to the next classification. The time to the next classification is a feature of the classification history rather than of a forecast. I.e., at each step it records the interval to the following classification, so it is defined only for steps that have a successor. It thus uses no information beyond what the classification history already provides, and the final, most recent classification simply carries no value for this feature.
These temporal features are normalized to the range $[0,1]$ and directly concatenated to the existing time series. 
As suggested by \cite{shukla2021multi}, concatenation-based time encodings have been widely adopted for modeling irregularly sampled time series \citep{zhang2019attain, song2018attend, tan2020data}, in contrast to directly adding positional encodings to input embeddings as used in standard transformers \citep{vaswani2017attention}. 

There are more sophisticated approaches for handling the irregular time series using learnable time embeddings or interpolations; 
additional discussion of this matter will be presented in Section \ref{sec:methods}. 
The resulting data take the form of a ten-dimensional multivariate time series, with five channels of classification probabilities, one channel for the raw flux, and four temporal feature channels at each timestamp. 
We use these multivariate time series directly without any further feature extraction or augmentation, aiming to retain the full raw information without complicating the framework. 

\section{Methodology}\label{sec:methods}

In this section, we present the main model architecture designed to incorporate the historical classifications and raw flux data. 
We discuss the motivation for our choices and potential alternative approaches. 
We then introduce the new metrics and assessment tools we propose for model evaluation, including relevant background on Wasserstein distances and visualization choices.

\subsection{Cascade Generalization}\label{sec:model}

We adopt a modeling approach similar to the cascade generalization \citep{gama2000cascade}, which can be considered a special case of a stacking algorithm \citep{wolpert1992stacked}. 
Unlike the standard stacking approach that combines base classifiers trained in parallel, the cascade generalization combines models in sequential order using an iterative composition to a classification pipeline. 
Under this set-up, each classifier receives the original input and the predicted probability distributions from a previous classifier as new features. 
In \cite{gama2000cascade}, such an approach was characterized as loose coupling, where the base classifier(s) pre-process the data as a feature extractor for the meta-classifier in the following stage. 
The main intuition behind such a framework is that ``the final model uses the representational language of the high-level classifier, possibly enriched with expressions in the representational language of the low-level classifiers'' \citep{gama2000cascade}. 
For instance, the base model may be a tree-based classifier such as a boosted decision tree, while the meta-classifier adopts a neural network-based model. 
This architectural diversity enables the two models to complement each other, yielding improved performance. 

We also propose an additional rationale for how such sequential coupling improves classification performance. 
In our case, the classification histories can be considered as soft labels that carry uncertainty-annotated suggestions. 
The meta-classifier can recalibrate the base classifications by rectifying systematic classification bias and errors by learning mistake patterns of the base model and its overconfidence/underconfidence behavior throughout the classification histories. 
A similar error correction idea was outlined in the original stacked generalization paper \citep{wolpert1992stacked}, where they ``use stacked generalization to improve a single generalizer.'' 
The outlined approach tries to estimate the prediction error based on the test point and its nearest neighbor from the training set. 
The estimated error (or a fraction of it) is added back to the actual prediction to obtain a corrected final result. 
In addition, by producing new classifications conditioned on previous classification histories, the meta-classifier naturally resembles an incremental learning style by taking into account both recent and earlier classifications. 
This could achieve robustness against short-term noise and flux fluctuations, whether caused by intrinsic variability of astronomical objects or observational uncertainties, resulting in a more stable classification.

\cite{gama2000cascade} also provides suggestions on the optimal strategy of combining classifier results by empirically comparing different coupling strategies of linear discriminant, C4.5 decision tree, and naive Bayes classifier evaluated with 26 data sets from the UCI repository \citep{asuncion2007uci}. 
They suggest combining classifiers with different bias-variance behaviors: low-level models use algorithms with low variance (simpler models like decision trees, which are relatively more consistent and stable but may miss complex patterns), while high-level models use algorithms with low bias (which capture more complex patterns but tend to overfit to training samples, like neural networks).
The underlying intuition is that lower-level learners defer final decisions to higher-level learners. 
By selecting low-bias learners for the higher level, we can fit more complex decision surfaces while leveraging the ``stable'' surfaces drawn by the low-level learners. 
Note that we have limited details related to the model architecture of the baseline classifiers in ELAsTiCC2, since alert brokers may adopt and implement different versions of the classifier for the challenge and for the public. 
Nevertheless, such an argument justifies our choice of using a neural network-based model for the meta-classifier, which is typically characterized as having high variance and low bias. 
A potential follow-up analysis is to include more details related to the baseline classifiers' architecture for a more complete evaluation of the coupling strategy. 

For our model, we treat the classification histories from existing classifiers as the lower-level inputs that extend the raw light curves to build a second-level classifier. 
Note that the classification histories provide a richer context that characterizes the temporal behavior of the base models. 
While we use only a single meta-classifier built upon one base model (a selected ELAsTiCC classifier), the cascade generalization framework can combine multiple models in a sequence.

\subsection{Model Architecture}\label{sec:model}

We propose a three-stage model that combines a recurrent network and attention mechanisms. Let the raw input be $\mathbf{X} \in \mathbb{R}^{B \times T \times C}$, where $B$ is the batch size, $T$ is the sequence length, and $C=6$ is the number of input feature channels (classification histories for five classes and one channel for normalized flux). When timestamp encoding is enabled, four additional temporal features are computed and concatenated with the original features at each timestep to produce the augmented input $\tilde{\mathbf{X}} \in \mathbb{R}^{B \times T \times (C + 4)}$.

In the first stage, variable-length time series of raw flux, classification PMFs, and timestamp encodings (if enabled) are preprocessed through a batch collation process. 
For each batch, the padding length is determined as the minimum of either the longest sequence in the batch or a predefined maximum of 100. 
Sequences exceeding this maximum are truncated by retaining only the first 100 observations, while shorter sequences are zero-padded to match the batch's target length. 
The original sequence lengths are preserved and passed to the LSTM \citep{hochreiter1997long} layer, enabling the use of PyTorch's \texttt{pack\_padded\_sequence} functionality. 
This approach ensures that padded values are excluded from LSTM computations, preventing the model from learning spurious patterns from the zero-padding and increasing the computation efficiency. 

After the pre-processing, the inputs are fed into a single-layer LSTM with the standard implementation by PyTorch \citep{paszke2019pytorch}:
\begin{equation}
    \mathbf{h}_t, \mathbf{c}_t = \text{LSTM}\!\left(\tilde{\mathbf{x}}_t,\, \mathbf{h}_{t-1},\, \mathbf{c}_{t-1}\right),
\end{equation}
where at each timestep $t$, the LSTM updates its hidden state $\mathbf{h}_t \in \mathbb{R}^{d_h}$ and cell state $\mathbf{c}_t \in \mathbb{R}^{d_h}$ from the previous states 
$\mathbf{h}_{t-1}$ and $\mathbf{c}_{t-1}$ and the current input $\tilde{\mathbf{x}}_t$ 
through gating mechanisms \citep{hochreiter1997long}. This yields hidden states $\mathbf{H} = [\mathbf{h}_1, \ldots, \mathbf{h}_T] \in \mathbb{R}^{B \times T \times d_h}$ that encode temporal dependencies across the observations, where $d_h = 256$.

Then, a binary padding mask is constructed from the sequence lengths $\{\ell_i\}$ for the $i$th sequence. At each timestep $t$:
\begin{equation}
    m_t^{(i)} =
    \begin{cases}
        1 & \text{if } t \leq \ell_i \\
        0 & \text{otherwise}
    \end{cases}
\end{equation}

The LSTM outputs $\mathbf{H}$ are masked element-wise as $\hat{\mathbf{h}}_t = \mathbf{h}_t \cdot m_t$ to zero out padded positions prior to any attention computation.

For the second stage, we apply an additive attention mechanism to the sequence of hidden states, allowing the model to selectively weight the importance of different time steps and aggregate them for the final classifications.
Instead of using the more standard dot-product attention \cite{vaswani2017attention}, we use an additive attention mechanism (Bahdanau attention; \cite{bahdanau2014neural}), which computes the compatibility function using a single-layer feed-forward network and a nonlinear $\tanh$ activation function. 
Though this sacrifices computational efficiency, the main goal is to capture complex nonlinear relationships among hidden states. This can be formulated as follow:

Attention scores are computed following an additive formulation \citep{bahdanau2014neural} without an external query, treating the mechanism as a learned content-based weighting LSTM hidden states $\mathbf{h}_t$ over timesteps. For each timestep $t$, an intermediate attention representation is obtained by:

\begin{equation}
    \mathbf{u}_t = \tanh\!\left(\mathbf{W}_a\, \hat{\mathbf{h}}_t + \mathbf{b}_a\right), \quad \mathbf{u}_t \in \mathbb{R}^{d_a}
\end{equation}

where $\mathbf{W}_a \in \mathbb{R}^{d_a \times d_h}$ is a learned weight matrix, $\mathbf{b}_a \in \mathbb{R}^{d_a}$ is a bias term, and $d_a = 64$ is the attention hidden dimension. A scalar attention score is then obtained via a learned projection vector $\mathbf{v} \in \mathbb{R}^{d_a}$:

\begin{equation}
    e_t = \mathbf{v}^\top {\mathbf{u}}_t
\end{equation}

Padded positions are excluded from normalization by assigning $e_t = -\infty$ where $m_t = 0$, ensuring they receive zero weight after the softmax. The attention weights are then computed via softmax function as:

\begin{equation}
    \alpha_t = \frac{\exp(e_t)}{\sum_{t'=1}^{T} \exp(e_{t'})}, \qquad \sum_{t=1}^{T} \alpha_t = 1
\end{equation}

The final attended representation is the weighted sum of the masked LSTM outputs (hidden states):

\begin{equation}
    \mathbf{z} = \sum_{t=1}^{T} \alpha_t\, \hat{\mathbf{h}}_t \in \mathbb{R}^{d_h}
\end{equation}

Finally, the attended representation $\mathbf{z}$ is fed into a three-layer fully connected network with hidden dimensions $[128, 64, 32]$ and output dimension $K = 5$, where batch normalization, ReLU activation, and dropout ($p = 0.4$) are applied after each of the first two layers.

The model offers a straightforward architecture that effectively incorporates the temporal information.

While the combination of standard LSTM and an additive attention mechanism does not represent the state-of-the-art in the handling irregular time series when compared to modern transformer-based architectures, the primary goal of our work is not to propose a novel model architecture, but rather to demonstrate the value of incorporating historical classifications into light curve classification using a straightforward architecture. 
More sophisticated approaches have been developed for irregular temporal data, including using neural-controlled differential equations for interpolation \citep{kidger2020neural}, modified LSTM architectures designed for irregularly-sampled sequences \citep{zhang2019attain}, transformers based on improved position encodings\citep{foumani2024improving}, and learnable time embeddings for attention mechanisms \citep{shukla2021multi}. 
Future work will explore integrating our approach with these advanced architectures. 
Additionally, while standard LSTMs can be less computationally efficient than transformers for processing long sequences due to their recurrent structure, this limitation is not a concern in our application since we impose an upper bound of 100 timestamps on the sequence length.

\subsection{Naive Model}
To demonstrate the benefits of including classification histories, we also proposed a naive model for a more comprehensive comparison. Instead of processing the full time series with recurrent network and attention, the naive model directly use the final classification PMFs for each object as inputs, which are the most recent classification PMFs obtained with the complete light curves. 

The final classification PMFs are directly fed into the classification head of the new model proposed above, which is a three-layer fully connected network with hidden dimensions $[128, 64, 32]$ and output dimension $K = 5$, where batch normalization, ReLU activation, and dropout ($p = 0.4$) are applied after each of the first two layers. Note that no classification histories or raw flux observations are used for the naive model training.

Our framework treats the selected ELAsTiCC2 classifiers as fixed base models and supplements the raw light curves with their classification histories. Most classifiers are trained primarily on complete or near-complete light curves and are not directly optimized for early classification using partial observations or truncated light curves. Therefore, part of the early instability and bias we observe in the baseline models, as measured by the early-stable classification metrics, may reflect this distribution shift.

Our meta-classifier improves performance in part by learning to correct such systematic mistakes of the base model, but this means our reported gains are measured relative to base models that may be suboptimal for early classifications. Another informative alternative naive model, complementary to the current choice, would be a classifier trained on early and partially observed light curves with varying degrees of truncation. Comparing against such a early classifier would help separate the performance gains added by exploiting classification histories from the gains that could be obtained by adapting the underlying classifier to the partially observed light curves. However, we note that this comparison is not directly feasible within the present study, since we work only with the pre-trained models and do not have access to the base classifiers or their training pipelines; constructing it would require either retraining brokers' classifiers on truncated data or developing a new classifier from raw light curves, which we leave to future work.

\subsection{New Early-Stable Classification Metrics and Assessment Tools}\label{sec:newmetric}

As motivated in Section \ref{sec:intro}, in addition to an accurate final classification, an ideal classifier would consistently assign high classification probabilities for the true class at earlier stages of the observations. 
In this subsection, we propose Early-Stable Classification Metrics that incorporate both a measure of early classification performance and classification stability via the Wasserstein distance from optimal transport theory. 
We will outline two versions of the metric and propose several visualizations for using the metric and for evaluating model stability in general.

\subsubsection{Wasserstein Distance}

An ideal evaluation methodology would account for three aspects of a classifier: accuracy, stability, and early performance. 
Classification accuracy can naturally be quantified through the classification PMFs, where higher classification probabilities assigned to the true class indicate better model performance. 
Early performance can be quantified by the number of observations or the time the classifier takes to achieve a certain pre-determined threshold for classification probability assigned to the true class.

To quantify the classification stability, we propose using the Wasserstein distance between probability distributions, developed in optimal transport theory \citep{villani2021topics, kolouri2017optimal, peyre2019computational} and applied previously in the context of astronomical light curve classification, e.g. \cite{malz2025light}.

It is instructive to start with the case of continuous probability distributions $P$ and
$Q$, two univariate probability distributions on $\mathbb{R}$.

To measure the distance between two distributions, one can employ the framework of integral probability metrics (IPMs; \cite{muller1997integral}), defined as
\begin{equation}
    d_{\mathcal{F}}(P, Q) = \sup_{f \in \mathcal{F}} \left| \mathbb{E}_{X \sim P}[f(X)] - \mathbb{E}_{Y \sim Q}[f(Y)] \right|,
\end{equation}
where $\mathcal{F}$ is a class of functions.

By restricting $\mathcal{F}$ to the class of 1-Lipschitz functions,
\begin{equation}
    \mathcal{F}_L = \{ f : |f(x) - f(y)| \leq \|x - y\|, \; \forall x, y \},
\end{equation}

we obtained the Wasserstein-1 distance:
\begin{equation}
    W_1(P, Q) = \sup_{f \in \mathcal{F}_L} \left| \mathbb{E}_{X \sim P}[f(X)] - \mathbb{E}_{Y \sim Q}[f(Y)] \right|.
\end{equation}

By Brenier's theorem \citep{brenier1991polar}, this is equivalent to an optimal transport problem:
\begin{equation}
   W_1(P, Q) = \min_T \mathbb{E}[\|T(X) - X\|] ,
\end{equation}
where the minimum is taken over all transport maps $T$ such that $T(X) \sim Q$ when $X \sim P$.

In our case, we need to compute the distance between two discrete distributions characterized by their probability mass functions. 
Compared with the continuous case described above, the Wasserstein distance in this case can be more intuitively understood as the earth mover's distance (EMD),
in which an analogy is made with the work associated with ``shifting'' one probability mass function into another, wherein each is imagined as a pile of earth. 

This can be formally defined as an optimization problem, as follows: 
For two discrete distributions $P$ and $Q$ with finite supports $\{x_1,\ldots,x_m\}$ and $\{y_1,\ldots,y_n\}$ respectively, let $\mathbf{p} = (p_1,\ldots,p_m)$ and $\mathbf{q} = (q_1,\ldots,q_n)$ denote the probability mass vectors where $p_i = P(x_i)$ and $q_j = Q(y_j)$, with $\sum_{i=1}^m p_i = \sum_{j=1}^n q_j = 1$.
Then
\begin{equation}\label{eq:5}
    W_1(P, Q) = \min_{\mathbf{T} \in \mathbb{R}_+^{m \times n}} \sum_{i,j} \mathbf{T}_{i,j} \mathbf{C}_{i,j},\\
\end{equation}
\begin{equation}\label{eq:6}
    \mbox{such that} \quad \mathbf{T} \mathbf{1} = \mathbf{p}; \quad \mathbf{T}^\top \mathbf{1} = \mathbf{q}; \quad \mathbf{T} \geq 0,
\end{equation}
where
\begin{itemize}
\item $\mathbf{C} \in \mathbb{R}_+^{m \times n}$ is the metric cost matrix, where $C_{i,j}$ represents the cost to move a unit mass from source support $x_i$ to target support $y_j$

\item $\mathbf{T} \in \mathbb{R}_+^{m \times n}$ is the transport plan matrix that specifies how the mass from the source distribution transfer to the target distribution, where $\mathbf{T}_{i,j}$ represents the amount of probability mass transported from $x_i$ to $y_j$

\item $\mathbf{1} \in \mathbb{R}^n$ denotes unit vector of ones
\end{itemize}

\begin{figure}
\centering
\includegraphics[width= 1\linewidth, scale=1] 
{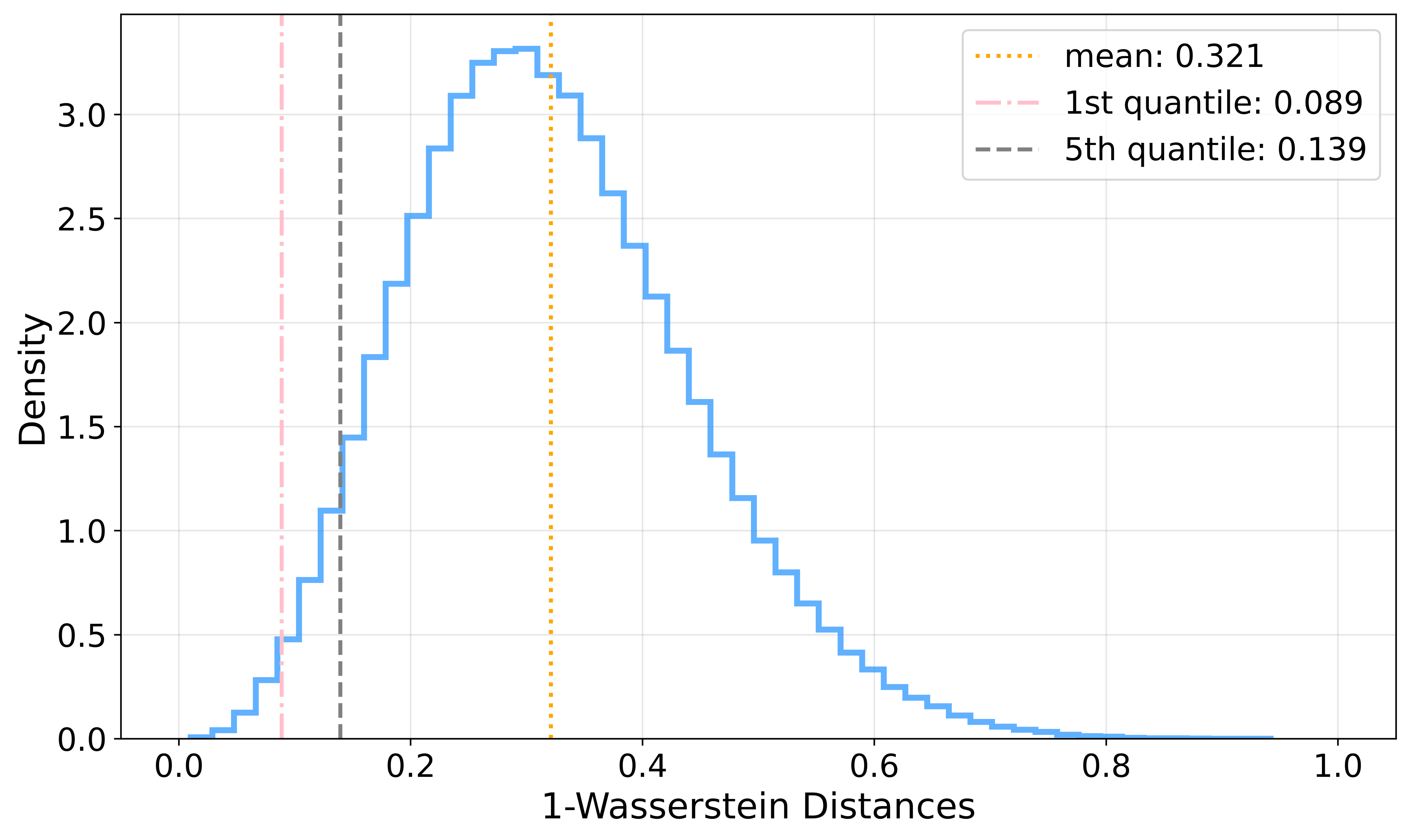} 
\caption{\label{fig:wasser} Histogram of 1-Wasserstein distances computed between 500,000 pairs of randomly simulated PMFs with five supports with mean (orange, dotted), 1st quantile (pink, dot-dashed), and 5th quantile (grey, dashed) plotted.
The distribution is slightly right-skewed.}
\end{figure}

The constraints ensure probability mass conservation after the transport, i.e., we must relocate all the mass from the source distribution to the target distribution. 
In our case, we have $n=m$ since we are comparing classification PMFs with the same supports, and both $\mathbf{T}$ and $\mathbf{C}$ are square matrices. 
Since supernova class labels don't have a natural ordering and hence cannot be used as valid supports, a reasonable choice is to manually define a unit cost matrix $\mathbf{C_u}$ with 0 on the diagonal and $\mathbf{C_u}_{i,j}=1,\ \forall i\neq j$. 
Notice that with the unit cost matrix $\mathbf{C}_u$, the 1-Wasserstein distance is bounded, $W_1(P, Q)\in[0,1]$, between all valid probability mass functions of finite support. 
In Figure \ref{fig:wasser}, we plot the histograms of 1-Wasserstein distances computed between 500,000 pairs of randomly generated PMFs with supports of 5. 
The distribution is slightly right-skewed with a mean around 0.32 and values between 0 and 1. 
The 1st and 5th quantiles of the simulated distribution are 0.089 and 0.139, respectively, which can serve as references when we define the metric later. 
Note that the cost matrix can be adjusted to reflect specific analysis objectives; for instance, lower costs may be assigned to changes in classification probability among more similar supernova classes.

The Wasserstein distance, formulated by Equation (\ref{eq:5}) and the constraint given in (\ref{eq:6}), can be computed using linear programming. 
In this work, we use the Python Optimal Transport library (POT; \citealt{flamary2021pot}) for the computation. 
Intuitively, a higher transport distance indicates a more drastic change in terms of the raw PMFs. 

In general, the Wasserstein distance can have several advantages, including but not limited to: (1) it is a valid distance metric between well-behaved probability distributions, (2) it accounts for the underlying geometry of the space, which can be ignored by measures like Kullback–Leibler (KL) divergence, (3) it can be simpler to estimate in practice via linear programming, (4) it is well-defined between a discrete and continuous distribution and between distributions with non-overlapping supports, which is a cornerstone property behind the Wasserstein Generative Adversarial Networks (WGAN; \cite{arjovsky2017wasserstein}). 

Here, a key advantage of the Wasserstein distance is its intuitive interpretation as the ``earth mover's distance'' and the associated transport matrix between distributions. 
The row sums of this matrix give the classification PMFs from the previous timestamp, and the column sums gives the PMFs for the current timestamps. 
The $(i,j)$ entry provides the probability mass moved from the class $i$ to class $j$, and the sum of the off-diagonal values gives the exact Wasserstein distance under the unit cost matrix $\mathbf{C}_u$. 
In this way, we not only quantify how two distributions differ, but also capture information on how classification PMFs evolve between observations. 
While not explored in this work, the transport matrix can provide additional insights into the temporal behavior of light curve classifiers. 
One potential usage of such information is to quantify and visualize whether the classifier is becoming more confident with more mass moved towards the true class and other more complicated classification dynamics.

\subsubsection{Early-Stable Classification Metrics}\label{sec:ecs}

We propose two metrics that jointly take into account classification accuracy, stability, and early classification performance using the Wasserstein distance as formulated in the previous section for stability quantification. 
We name the two metrics Early-Stable Classification Fraction ($\ESCf$) and the Early-Stable Classification Score ($\ESCs$). 

We first define the Early-Stable Classification Fraction ($\ESCf$) as follows:

\begin{definition}[Early-Stable Classification Fraction]
\label{def:escf}
Given a sequence of classification PMFs from an object for a $n$-class classification task indexed by $t$, $\{p_t\}_{t=1}^T$, where $T$ is the total number of observations, the classification converges $(\epsilon, \rho, k)$-fast with convergence time $\tau$, $1\leq\tau\leq T$, if both of the following conditions are satisfied:

\begin{enumerate}
\item \textbf{Stability Condition}: There exist $k$ consecutive classifications, $k\leq T$ and $k\geq2$, up to and including time $\tau$ such that
$$W_1(p_t, p_{t+1}) \leq \epsilon \quad \forall t \in \{\tau-k+1, \tau-k+2, \ldots, \tau-1\},$$
for some fixed $\epsilon\in(0,1)$, and $W_1(\cdot, \cdot)$ is the Wasserstein-1 distance computed for the two PMFs with a cost matrix that is selected to have a unit cost of moving unit mass between any distinct classes.

\item \textbf{Accuracy Condition}: For some fixed $\rho \in [0,1]$,
$$p_{t,*}\geq \rho \quad \forall i \in \{\tau-k+1, \tau-k+2, \ldots, \tau\},$$
where $p_{t,*}$ denotes the classification probability for the true class at time $t$. 
\end{enumerate}

Then, we define the Early-Stable Classification Fraction ($\ESCf$) for the object as:
\begin{equation}
    \ESCf = \begin{cases}
        \tau/T & \text{if converged} \\
        1 & \text{otherwise}
    \end{cases}.
\end{equation}

\end{definition}

Ideal classifiers have a higher proportion of objects that converged per class and a smaller convergence fraction per object. 
Based on the simulations in Figure \ref{fig:wasser}, choices for $\epsilon$ could be 0.05 (less than the 1st quantile) for a stricter requirement and 0.1 (between the 1st and 5th quantiles) for a relatively relaxed tolerance. 
The choice of $\rho$ and $k$ is more application-specific, with larger values corresponding to stricter requirements. 
Note that while we define convergence time by the number of observations, one could also use other measures, such as the elapsed time since first detection.

The above convergence-time version of the new metric offers a more intuitive understanding of the classifier's early-stable classification performance. 
For a more standardized and concise summary, we propose a weighted-sum version of the metric, named the Early-Stable Classification Score ($\ESCs$), defined as follows:

\begin{definition}[Early-Stable Classification Score]
\label{def:escs}

Given a sequence of classification PMFs for an $n$ class classification task indexed by $t$, $\{p_t\}_{t=1}^T$, where $T$ is the total number of observations, we define a weight curve (typically non-decreasing) $w(t):[0,1]\rightarrow [0,1]$. 
Then, we have:

\begin{equation}
    \ESCs=-\frac{1}{Z}\sum_{t=1}^{T-1} w\!\left(\frac{t}{T}\right) W_1(p_t, p_{t+1}) \ln(p_{t,*}),
\end{equation}
where $Z=\sum_{t=1}^{T-1}w\!\left(\frac{t}{T}\right) W_1(p_t, p_{t+1})$ is the sum of weights for normalization and $p_{t,*}$ is the classification probability for the true class at time $t$.

\end{definition}
Each metric variant offers distinct trade-offs, making them suitable for different application scenarios. 

\subsubsection{Application of $\ESCf$}
The fraction metric provides a measure of a model's early classification performance by combining early classification performance, stability, and accuracy into a single score. 
Users can flexibly adjust the stability parameters ($\epsilon, k$) and accuracy threshold ($\rho$) based on specific research objectives. 

For example, applications requiring high-confidence classifications to avoid wasting observational resources on false positives can employ stringent criteria such as a high accuracy threshold of $\rho\geq0.9$ with very small stability tolerance of $\epsilon=0.01$ with $k=10$. 
On the other hand, for rare transients with limited observations, one can allow more uncertainty and apply more permissive thresholds by lowering the accuracy threshold to $\rho\geq0.5$ with medium stability tolerance of $\epsilon=0.1$ with $k=5$. 
By expressing convergence time as a proportion $\tau/T$ instead of absolute counts, the metric is bounded in $(0,1]$ across all objects and classes, where lower values indicate the classifier achieves accurate and stable classification at an earlier stage of the classification. 

The fraction metric enables several approaches and visualizations for model evaluation and comparison. 
For fixed stability and accuracy parameters $(\epsilon, \rho, k)$, we can compute the proportion of objects that achieve convergence within each class. 
For cross-classifier comparisons, we can calculate a weighted convergence proportion by aggregating class-level proportions weighted by class sizes, where higher values indicate better classification performance under the specified criteria. 

The classification fraction $\ESCf$ can be averaged within each class and/or aggregated across classes using a class-size weighted average for cross-model comparison. 
We can also use class-level histograms or density plots of convergence fractions to make a more direct comparison of classifiers for different object types. 
To evaluate model performance across various convergence criteria, we can construct two-dimensional heatmaps with different stability and accuracy requirements. 
For a fixed $k$, with the stability threshold $\epsilon$ and accuracy threshold $\rho$ on the x-axis and the y-axis, we can compute either the weighted convergence proportions or classification fraction $\ESCf$ for each cell across different combinations of $(\epsilon, \rho)\in[0,1]\times[0,1]$. 
The same heatmap approach applies to class-level comparisons by using class-level convergence proportions or average convergence fractions. 
We will use the weighted average for model comparison in section \ref{sec:new eval} and present some of these visualizations in section \ref{sec:visualization}.

\subsubsection{Application of $\ESCs$}

There are some potential limitations of the fraction metric. 
First, its discrete, threshold-based formulation makes it unsuitable as a differentiable loss function for gradient-based optimization during model training. 
Second, the binary nature of the accuracy threshold $\rho$ treats all classifications exceeding the threshold identically, disregarding finer distinctions in classification confidence. 
Third, while the flexibility to adjust parameters $(\epsilon, \rho, k)$ enables research-specific applications, this can hinder standardized comparisons across studies. 
These limitations motivate the use of the weighted sum variant. 

Instead of relying on two convergence conditions, $\ESCs$ incorporates early classification performance and classification stability as weights, $w\!\left(\frac{t}{T}\right)$ and $W_1(p_t, p_{t+1})$, which are directly applied to the (log) classification probabilities. 
Under a monotonically increasing weight curve $w(t)$, later and unstable classifications are assigned higher weights, leading to higher log probability loss. 
Objects with a smaller score have better early-stable classification performance.  

Despite being more standardized, this metric does offer customization options and flexibility through the weight curve $w(t)$. 
For instance, in supernova classification, the weight function could assign high penalties during the pre-explosion phase and near peak when accurate classification is critical for follow-up decisions, and lower penalties post-peak and at later post-decay times. 
Similar to the fraction variant, we can compute the average score per class and/or aggregate the class averages across classes using a class-size weighted average. 

With the overall sum structure and differentiability of the 1-Wasserstein distance, the weighted-sum version could be further adapted to a loss function for model training. 
A notable example that incorporates the Wasserstein distance in the loss function for model optimization is the Wasserstein Generative Adversarial Networks (WGANs; \cite{arjovsky2017wasserstein}), which minimized an approximation of the earth mover's distance. 
Here, instead of sticking to a standard choice such as cross-entropy loss that mainly focuses on the final classifications' precision-recall performance, one can adapt $\ESCs$ as the loss function for the gradient-based optimization. 
Such potential adaptations can expand the role of the proposed metric beyond model evaluation and may lead to improvements in the model for stable and early classifications more generally, which represents a promising direction for future work. 

\section{Experiments and Results}\label{sec:experiments}

In this section, we make a holistic evaluation of the proposed approach by comparing it with the baseline classifiers and a group of naive classifiers whose inputs are only the final classification PMFs instead of the full histories. We first evaluate the classifiers with standard metrics, including accuracy, weighted precision, F1 score, and a confusion matrix

Then, we apply the new metrics proposed in Section \ref{sec:newmetric} to assess models' stability and early performance. 
We also give examples of some useful visualizations based on the new metrics.

\subsection{Model Implementation}
The proposed models (include the naive model) in Section \ref{sec:model} are implemented with PyTorch and trained with a standard cross-entropy loss and an Adam optimizer \cite{paszke2019pytorch, kingma2014adam}. 
We fixed the choices of hyperparameters of the model instead of performing cross-validation for hyperparameter tuning. 
We use a 256 hidden state dimension for the single-layer LSTM with the standard implementation by PyTorch \citep{paszke2019pytorch} and a 64 dimension for the additive attention module. 
A dropout of 0.4 is applied to the tanh-activated attention hidden states prior to score projection (the final linear mapping to a scalar score), serving as a regularizer over the attention computation.
We apply a 70-30 train-test split, and 30\% of the training set is held out as the validation set. 
Models were trained for up to 200 epochs with early stopping (with patience equal to 20) based on classification accuracy on the validation set. 
The final model was selected to be the one with the highest validation accuracy and evaluated on the test set for comparison against the baseline model.  

Note that all evaluations based on classical metrics are computed using the final classification PMFs for each object, which are the most recent classification PMFs obtained with the complete light curves. For the confusion matrices, the final label are assigned to the class with highest classification probability from the final classification PMFs.

To ensure a more robust comparison across all three classifier types, we repeated the training process ten times using different random seeds for the train-test split. 
All reported evaluation metrics and confusion matrices represent the mean values across these ten iterations, with error bars indicating $\pm$ 1 standard deviation. 

\subsection{Evaluation with Accuracy, Precision, and F1 Score }\label{sec:eval classic}

\begin{table*}
  \caption{Comparisons of accuracy, weighted precision, and weighted F1 score}
  \centering
  \setlength{\tabcolsep}{4pt}
  \begin{tabular}{lc|ccc|c}
    \toprule
    & & \multicolumn{3}{c|}{\textbf{Models}} & {\textbf{Improvements}} \\
    \cmidrule(lr){3-6}
    \textbf{Metric} & \textbf{Classifier} & \textbf{Baseline} & \textbf{Naive} & \textbf{New} & \textbf{New - Baseline}\\
    \midrule
    \multirow{3}{*}{Accuracy} 
    & A & 0.894 $\pm$ 0.00059 & 0.902 $\pm$ 0.0008 & \textbf{0.913 $\pm$ 0.0015} & + 1.9\% \\
    & B & 0.867 $\pm$ 0.0017 & 0.902 $\pm$ 0.00088 & \textbf{0.914 $\pm$ 0.00098} & + 4.5\%\\
    & C & 0.751 $\pm$ 0.0014 & 0.805 $\pm$ 0.0014 & \textbf{0.871 $\pm$ 0.0012} & + 12\% \\
    \midrule
    \multirow{3}{*}{\shortstack[l]{Weighted\\Precision}} 
    & A & 0.891 $\pm$ 0.00071 & 0.900 $\pm$ 0.00088 & \textbf{0.911 $\pm$ 0.0017} & + 2\%\\
    & B & 0.887 $\pm$ 0.0013 & 0.899 $\pm$ 0.00085 & \textbf{0.913 $\pm$ 0.001} & + 2.7\%\\
    & C & 0.820 $\pm$ 0.0013 & 0.800 $\pm$ 0.0013 & \textbf{0.869 $\pm$ 0.0013} & + 4.9\%\\
    \midrule
    \multirow{3}{*}{\shortstack[l]{Weighted\\F1 Score}} 
    & A & 0.888 $\pm$ 0.00058 & 0.899 $\pm$ 0.00072 & \textbf{0.911 $\pm$ 0.0015} & + 2.3\%\\
    & B & 0.874 $\pm$ 0.0015 & 0.899 $\pm$ 0.00065 & \textbf{0.913 $\pm$ 0.001} & + 3.9\% \\
    & C & 0.777 $\pm$ 0.0013 & 0.800 $\pm$ 0.0013 & \textbf{0.868 $\pm$ 0.0013} & + 9.1\%\\
    
    \bottomrule
  \end{tabular}
  \label{table-classic}
\end{table*}

We begin by evaluating classifiers using classical metrics, including accuracy, precision-recall, F1 scores, and confusion matrices. 
Table \ref{table-classic} presents the overall classification accuracy, support-weighted precision, and F1 scores for the three selected ELAsTiCC classifiers. 
Figure \ref{fig5} displays the confusion matrices for classifiers A and C across the baseline, naive, and proposed classifiers. 
The confusion matrices for classifier B are provided in the appendix. 
We omit discussion of its confusion matrices, as this classifier originates from the same alert broker as classifier A and exhibits similar behavior.

In Table \ref{table-classic}, the newly proposed classifier outperforms both the baseline and naive classifiers in chosen metrics across all selected ELAsTiCC classifiers. 
In terms of overall accuracy, for classifiers A and B, the proposed method achieves accuracy improvements of approximately 2\% and 4.5\% over the baseline, respectively, with accuracies reaching 0.913 and 0.914. 
The new model shows a more substantial improvement in accuracy on classifier C, with accuracy increasing from 0.751 (baseline) to 0.871 (proposed), representing a gain of approximately 12 percentage points. 
Notably, baseline classifiers A and B already achieve nearly 90\% accuracy, suggesting limited room for improvement. 
This indicates that we are working at the performance frontier with well-built models, where even marginal improvements of a few percentage points represent valuable advances. 
In contrast, baseline classifier C exhibits relatively poor performance at approximately 75\% overall accuracy, leaving greater potential for enhancement.

We note that the naive classifiers also made improvements over the baseline, but are still outperformed by the new classifiers. 
Similar trends are observed in weighted precision and F1 scores, where the naive approach shows intermediate performance (except for the weighted precision for classifier C) and the proposed classifiers demonstrate superior performance across all metrics. The standard deviations across the ten iterations remain consistently small across all metrics.

\subsection{Evaluation with Confusion Matrix}\label{sec:eval classic}
\begin{figure*}
  \centering
  \includegraphics[scale=0.45]{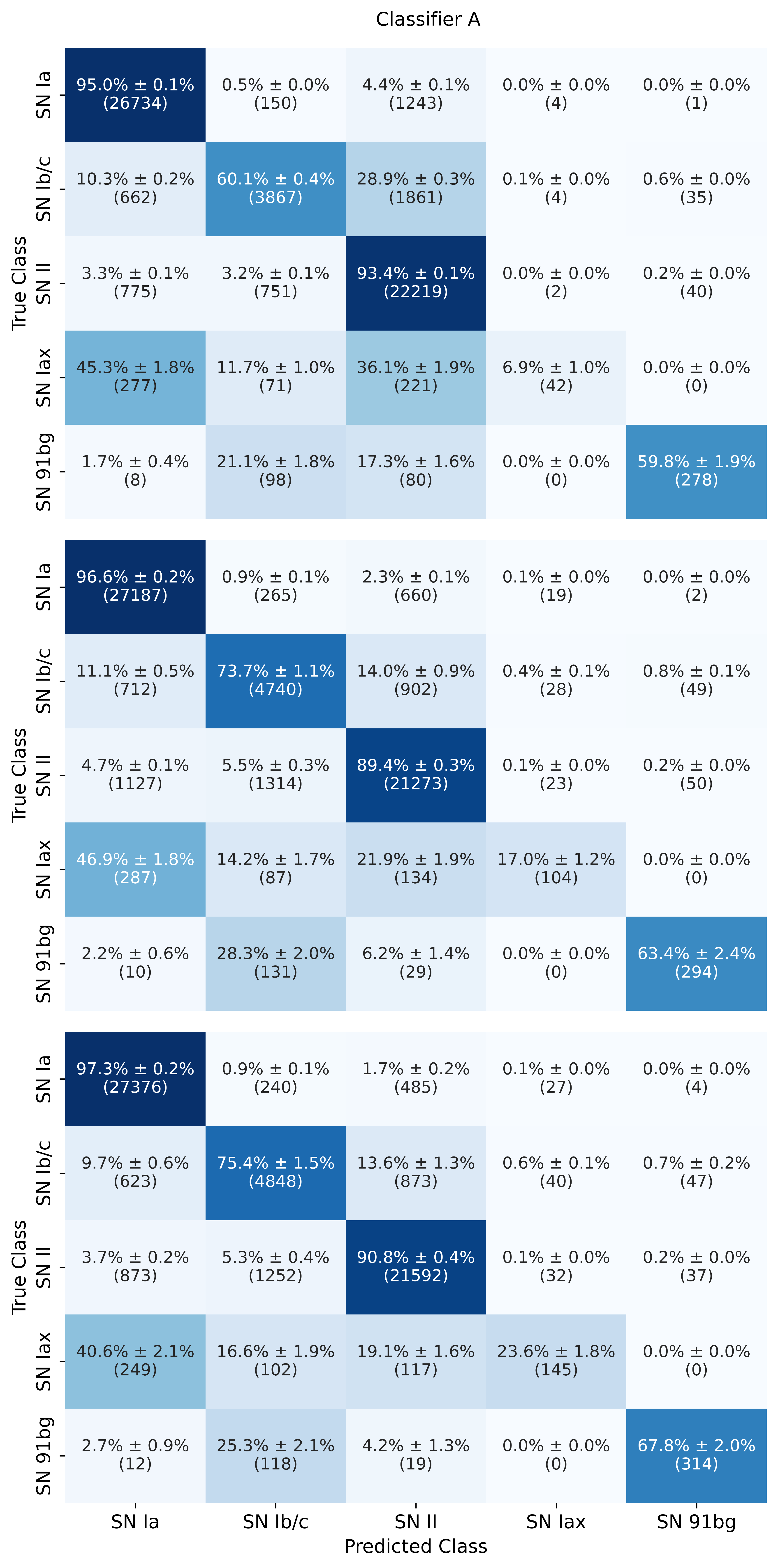} 
  \includegraphics[scale=0.45]{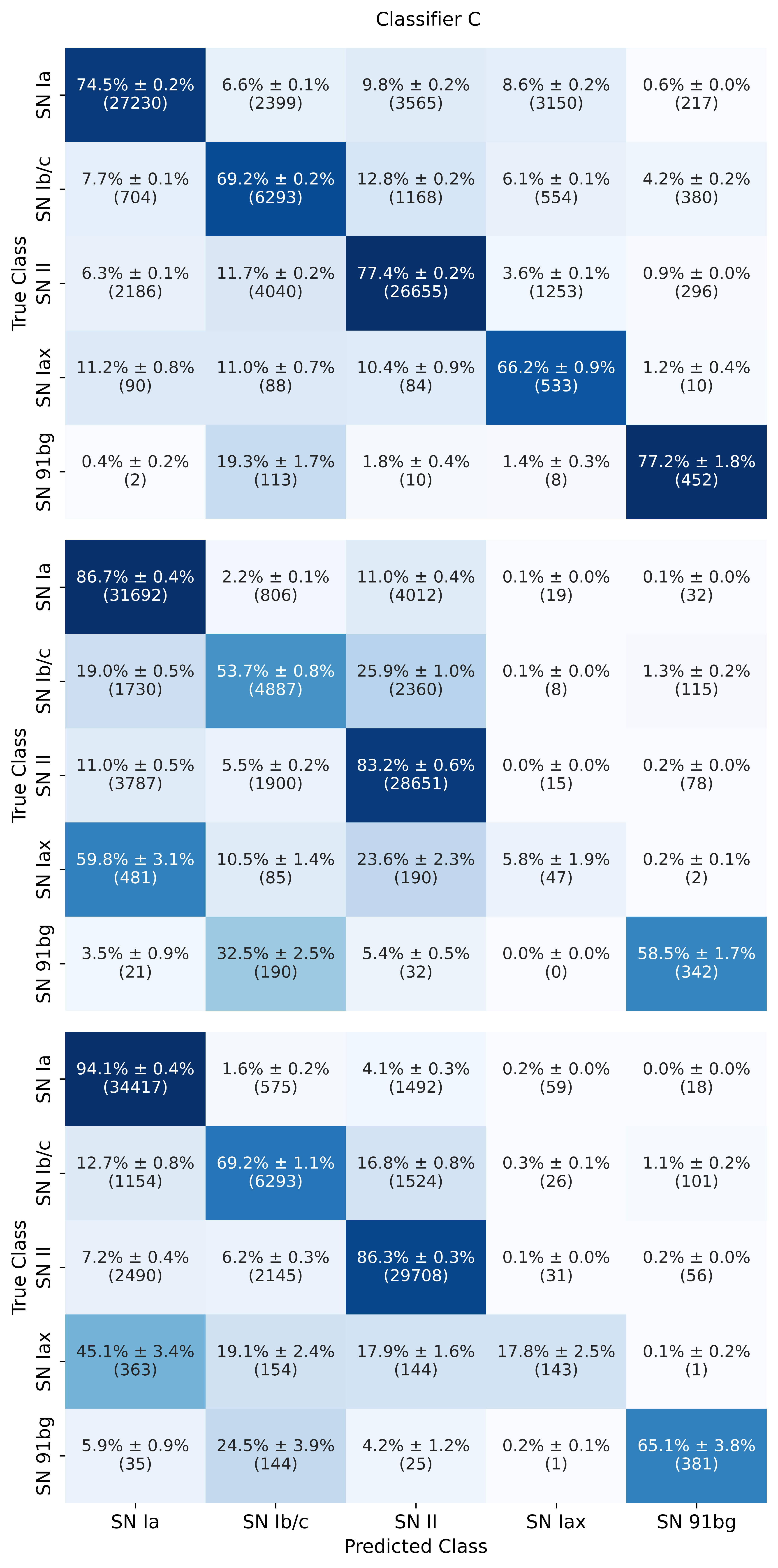} 
  \caption{The comparisons between the baseline (upper row), naive (middle), and new (bottom row) models across classifiers A (left) and C (right). 
  The confusion matrix is normalized per row and annotated with average absolute counts.
  The new models show improvements in overall accuracy and more balanced precision-recall.}
  \label{fig5}
\end{figure*}

We next examine per-class performance through confusion matrices presented in Figure \ref{fig5}. Note that we have three majority classes (SN Ia, SN Ib/c, and SN II) and two minority classes (SN Iax and SN 91bg) with significantly smaller sample sizes. 
More details of the sample sizes can be found in section \ref{sec:data}.

For the baseline classifier A, we observe high recall for SN Ia and SN II, with relatively lower recall for SN Ib/c and SN 91bg, and substantially poorer performance for SN Iax. 
While the two minority classes achieve relatively high precision, their recall remains low. 
The proposed model improves recall across all classes, with only a small reduction for SN II. 
However, such a reduction does not imply a worse performance of the new classifier. 
To be more specific, despite achieving a higher overall recall, the baseline classifier A exhibits a substantial false positive rate for SN II class, misclassifying 28.9\% of SN Ia objects and 4.4\% of SN Ib/c objects as SN II, which are the major misclassifications for both majority classes. 
This can be considered a systematic bias in the baseline classifier, where the classifier more easily (mis)classifies objects as SN II objects. 
Such poor precision poses practical concerns, potentially leading to misallocation of observational resources and undermining the model's reliability in real-world applications. 
The proposed model effectively addresses these limitations, achieving an improved recall-precision balance that enhances overall recall without substantial precision losses across any class. 
Nevertheless, we should note that the recalls for the two minority classes, particularly SN Iax, are still relatively low. 
This may be attributed to the substantially smaller sample sizes for these classes, which pose challenges for building satisfactory classifiers. 
The naive model shows similar intermediate improvements but is still outperformed by the new classifier. 

Baseline classifier C exhibits relatively poorer baseline performance compared to classifiers A and B. 
The recall is more uniform across all classes, including the two minority classes. 
The proposed model improves or maintains recall for all majority classes (with SN Ib/c remaining unchanged), achieving gains of approximately 20 and 10 percentage points for SN Ia and SN II, respectively, while maintaining satisfactory precision. 
In contrast, recall for both minority classes decreases under the proposed model. 
However, this follows the same recall-precision trade-off, analogous to the SN II behavior observed in classifier A. 
More specifically, while the baseline classifier C achieves 66.2\% recall (533 true positives) for SN Iax, the cost is 4,965 false positives. 
The high recall becomes less meaningful under such low precision. 
A similar pattern appears for SN 91bg, which has 452 true positives but incurs 903 false positives. 
The new classifier substantially reduces the number of false positives and achieves a more balanced precision-recall performance. 
As for the naive model, in terms of recall, it achieves a higher recall for class SN Ia and SN II but substantially sacrifices the recall for class SN Ib/c and SN Iax. 
This also suggests that the naive classifier may simply learn different trade-offs between classes, rebalancing performance rather than acquiring additional discriminative information to improve overall classification. 

In summary, the new classifiers achieve higher classification accuracy, weighted precision, and F1 scores across all selected ELAsTiCC classifiers compared with the corresponding baseline and naive classifiers. 
While the new models have reduced recall for certain classes, they achieve a worthwhile trade-off by gaining a more balanced precision-recall performance. 

\subsection{Evaluation with Early-Stable Classification Metrics}\label{sec:new eval}

\begin{table*}
  \caption{Comparisons of Early-Stable Classification Metrics and Average Changes}
  \centering
  \setlength{\tabcolsep}{4pt}
  \begin{tabular}{lc|ccc|c}
    \toprule
& & \multicolumn{3}{c|}{\textbf{Models}} & {\textbf{Improvements}} \\
    \cmidrule(lr){3-6}
    \textbf{Metric} & \textbf{Classifier} & \textbf{Baseline} & \textbf{Naive} & \textbf{New} & \textbf{New - Baseline} \\
    \midrule
    \multirow{3}{*}{\shortstack[l]{Convergence \\ Proportion}}
    & A & 0.836 $\pm$ 0.001 & 0.511 $\pm$ 0.0026 & \textbf{0.873 $\pm$ 0.0016} & + 3.7\% \\
    & B & 0.757 $\pm$ 0.0017 & 0.515 $\pm$ 0.0028 & \textbf{0.872 $\pm$ 0.0031} & + 11.5\% \\
    & C & 0.532 $\pm$ 0.0015 & 0.431 $\pm$ 0.0038 & \textbf{0.751 $\pm$ 0.0048 } & + 21.9\% \\
    \midrule
    \multirow{3}{*}{\shortstack[l]{Early-Stable\\ Classification Fraction\\(smaller values are preferable)}}
    & A & 0.475 $\pm$ 0.00071 & 0.684 $\pm$ 0.0037 & \textbf{0.451 $\pm$ 0.0076} & - 0.024 \\
    & B & 0.580 $\pm$ 0.001 & 0.678 $\pm$ 0.0043 & \textbf{0.474 $\pm$ 0.011} & - 0.106\\
    & C & 0.776 $\pm$ 0.00092 & 0.793 $\pm$ 0.0032 & \textbf{0.655 $\pm$ 0.0093} & - 0.121\\
    \midrule
    \multirow{3}{*}{\shortstack[l]{Early-Stable\\ Classification Score \\ (smaller values are preferable)}} 
    & A & \textbf{0.414 $\pm$ 0.0022} & 1.758 $\pm$ 0.031 & 0.441 $\pm$ 0.0074 & + 0.027\\
    & B & 0.540 $\pm$ 0.0031 & 1.853 $\pm$ 0.024 & \textbf{0.461 $\pm$ 0.015} & - 0.079\\
    & C & 0.900 $\pm$ 0.0029 & 1.407 $\pm$ 0.014 & \textbf{0.694 $\pm$ 0.030} & - 0.206\\
    \midrule
    \multirow{3}{*}{\shortstack[l]{Average\\Changes}} 
    & A & 1.874 $\pm$ 0.0062 & 1.435 $\pm$ 0.035 & \textbf{1.403 $\pm$ 0.051} & - 0.471\\
    & B & 2.867 $\pm$ 0.011 & 1.548 $\pm$ 0.12 & \textbf{1.509 $\pm$ 0.14} & - 1.358\\
    & C & 3.264 $\pm$ 0.0091 & 2.381 $\pm$ 0.06 & \textbf{2.069 $\pm$ 0.012} & - 1.195\\
    \bottomrule
  \end{tabular}
  \label{table-new}
\end{table*}

We next evaluate model performance using the Early-Stable Classification Metrics proposed in Section \ref{sec:ecs}, alongside some supplementary measures. 
The convergence parameters are fixed at less stringent settings with $\epsilon=0.1$, $\rho=0.5$, and $k=5$ for all evaluations. 
For baseline models, we directly evaluate existing historical classification PMFs. 
For the naive and new models, we construct classification histories of test objects through a sequential truncation procedure, as follows: 
For each time step $t\in\{2,\ldots, T\}$, we truncate the input time series of classification PMFs and raw flux to the first $t$ observations and apply the model to obtain the classification PMF at time $t$, which gives the complete sequence of classification PMFs after iterating through all $t$.

For each object class, we compute (1) the proportion of objects achieving convergence under the specified criteria, and (2) the average $\ESCf$ and $\ESCs$ values across all objects within the class. 
These class-level metrics are then aggregated via a weighted average, with weights proportional to class sample sizes. 
In addition, for each object, we also counted the number of times the classification label (class with the highest classification probability) changes throughout the classification histories as a more straightforward measure of the classification stability. 
We determine the average changes per class and aggregate them using the same weighted average for each classifier. 
We summarize the results in Table \ref{table-new}. 

As shown in Table \ref{table-new}, the proposed classifier achieves higher convergence proportions compared to both the baseline and naive classifiers. 
The improvements are 3.7\% and 11.5\% for classifiers A and B, respectively, and a substantial 21.9\% for classifier C, indicating that the proposed model yields more objects with stable classifications under the accuracy threshold $\rho=0.5$. 
Notably, the naive classifiers, which utilize only the most recent classification PMFs, can actually degrade classification stability, as suggested by the decreased convergence proportions across all three classifiers. 

Similar trends are observed in the Early-Stable Classification fraction. 
The proposed model achieves stable and accurate classifications at earlier stages of the observation series, as compared with the baseline and naive classifiers, as indicated by the smaller fractions across all three approaches. 
The improvement is relatively small for classifier A and more substantial for classifiers B and C. 
Again, the naive classifiers show worse performance when compared with the baseline, with higher fractions for all classifiers A, B, and C. 

The proposed classifier exhibits fewer changes in the classification label throughout the histories. 
The naive classifiers show smaller average changes with improved stability compared to the baseline models. 
However, we emphasize that the convergence proportion and the Early-Stable Classification Fraction we defined should be considered as a more comprehensive and stringent metric, as it requires a sequence of stable classifications and explicitly imposes a stability condition that directly quantifies variations in classification PMFs. 
The average changes should be used as a quick and intuitive summary instead of a more serious model evaluation. 

In regards to the Early-Stable Classification score, we use an exponential weight curve $w(t)$ on $[0,1]$ with adjustable steepness defined as
\begin{equation}
w(t; \alpha) = \frac{e^{\alpha t} - 1}{e^{\alpha} - 1}, \quad t \in [0,1], \quad \alpha > 0,
\end{equation}
where $\alpha$ is the steepness parameter, with higher values giving steeper curves that penalize more heavily on late classifications. 
We select $\alpha=2$ for our evaluations. 

While the proposed classifier outperforms both the baseline and naive classifiers for classifiers B and C, we observe that it performs slightly worse than classifier A. 
The new model has a higher Early-Stable Classification Score with a slight increase of 0.027 compared with the baseline model. 
We attribute this to several potential causes. 
First, the original classifier A already achieved a satisfactory overall classification performance with close to 90\% classification accuracy, and the improvement in other classic and new metrics is also relatively small compared with classifiers B and C. 
For instance, we only achieved 3.7\% and -0.024 improvement for the convergence proportion and Early-Stable Classification Fraction for classifier A, which are significantly smaller compared with those of classifier B and C. 
These suggest that we are refining predictions at the margins. 
We provide the detailed Early-Stable Classification Scores computed for each class for classifier A before aggregation in Appendix \ref{sec:appendix}. 
For the three majority classes, note that we achieved smaller or equal scores for SN Ia and SN Ib/c objects, but a worse performance with a larger score for SN II objects. 
This matches the previous discussion of the confusion matrix for classifier A, where we improved the recall for all other classes except SN II. 
We achieved a better overall precision-recall balance by trading a small reduction in SN II recall for improved precision. 
Note that the Early-Stable Classification Score presented in the table is computed as a class-size weighted average. 
Therefore, the reduced recall for SN II class, the second-largest class after SN Ia, leads to an increase in the overall score by reducing the average probability assigned to true classes. 
This also highlights the importance of examining detailed per-class metrics rather than relying solely on overall weighted summaries for a comprehensive assessment of model performance.

In total, the proposed classifier not only outperforms the baseline and naive classifiers in terms of classical metrics such as overall accuracy, weighted precision, and F1 score, but also achieves earlier and more stable classifications as evaluated with the proposed Early-Stable Classification Metrics (with one exception as mentioned above). 
Nevertheless, the magnitude of improvement varies across baseline classifiers, with more substantial gains for classifiers B and C but modest improvements for classifier A. 
The proposed early-stable classification metrics combined the evaluation of both early classification and classification stability. 
These two aspects are inherently coupled with the definition of our metrics, and improvement in one leads to an overall improvement of the metrics. 
Note that our new classifier's training objective remains the standard cross-entropy loss, without explicit optimization for early-stable classification. 

A more detailed analysis of the mechanism of why such modeling strategies improve the classification accuracy and early-stable classification performance can be an important direction for future work. 
For instance, one could start by examining the misclassified objects for both the baseline and the new classifier and analyzing the error correlation, which may provide a clearer view of how the new model makes improvements on previous misclassifications. 

\subsection{Comparisons of Classification PMFs}

We now examine the detailed classification PMFs produced by each classifier for some selected objects, providing additional insight into how the baseline, naive, and proposed classifiers differ in terms of their classification behavior at the object level. 
For this purpose, we focus on classifier A and present classification results for one SN Ia and one SN Ib/c object.

\begin{figure}
  \centering
  \includegraphics[scale=0.45, width=1\linewidth]{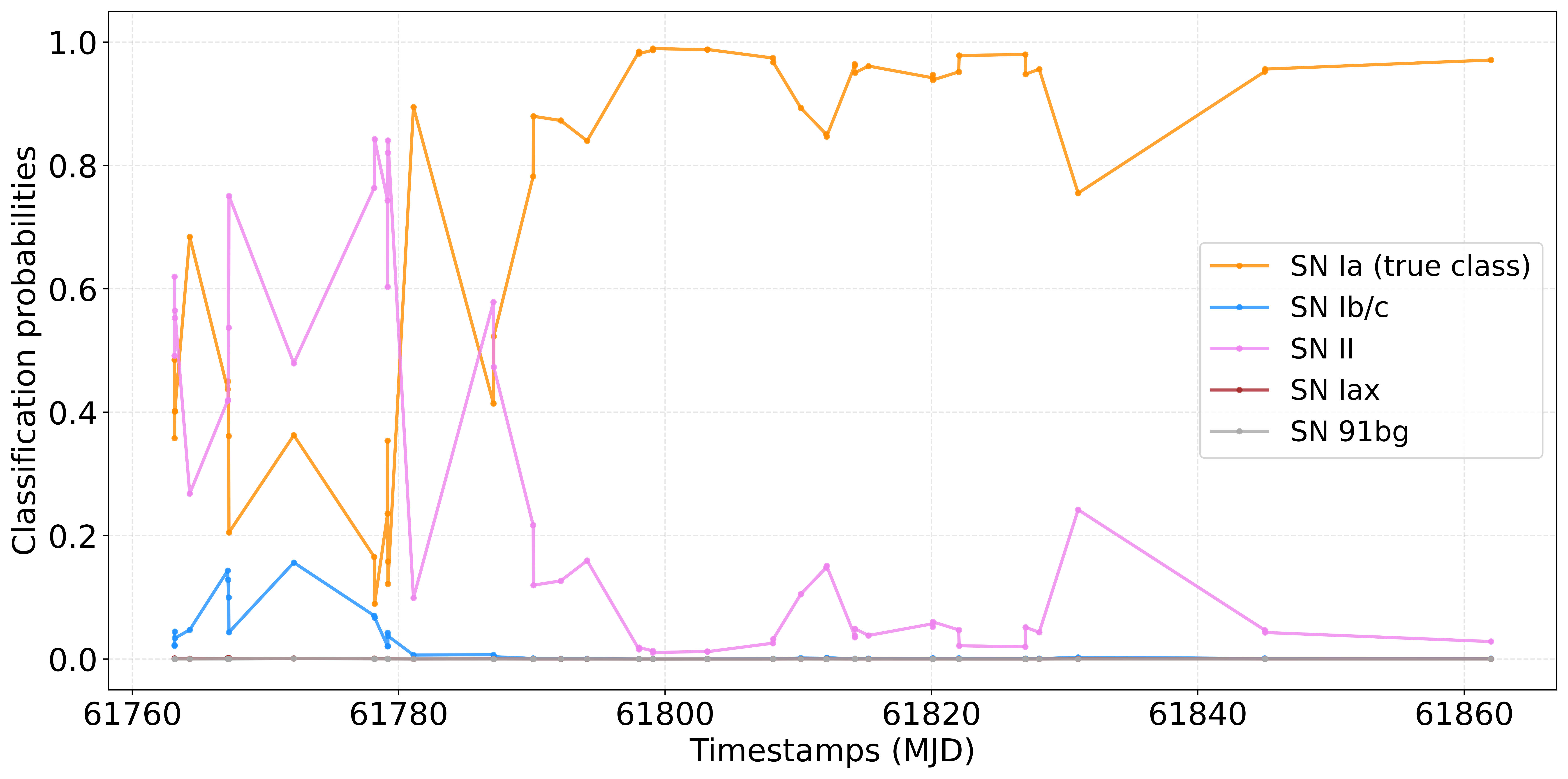} 
  \includegraphics[scale=0.45, width=1\linewidth]{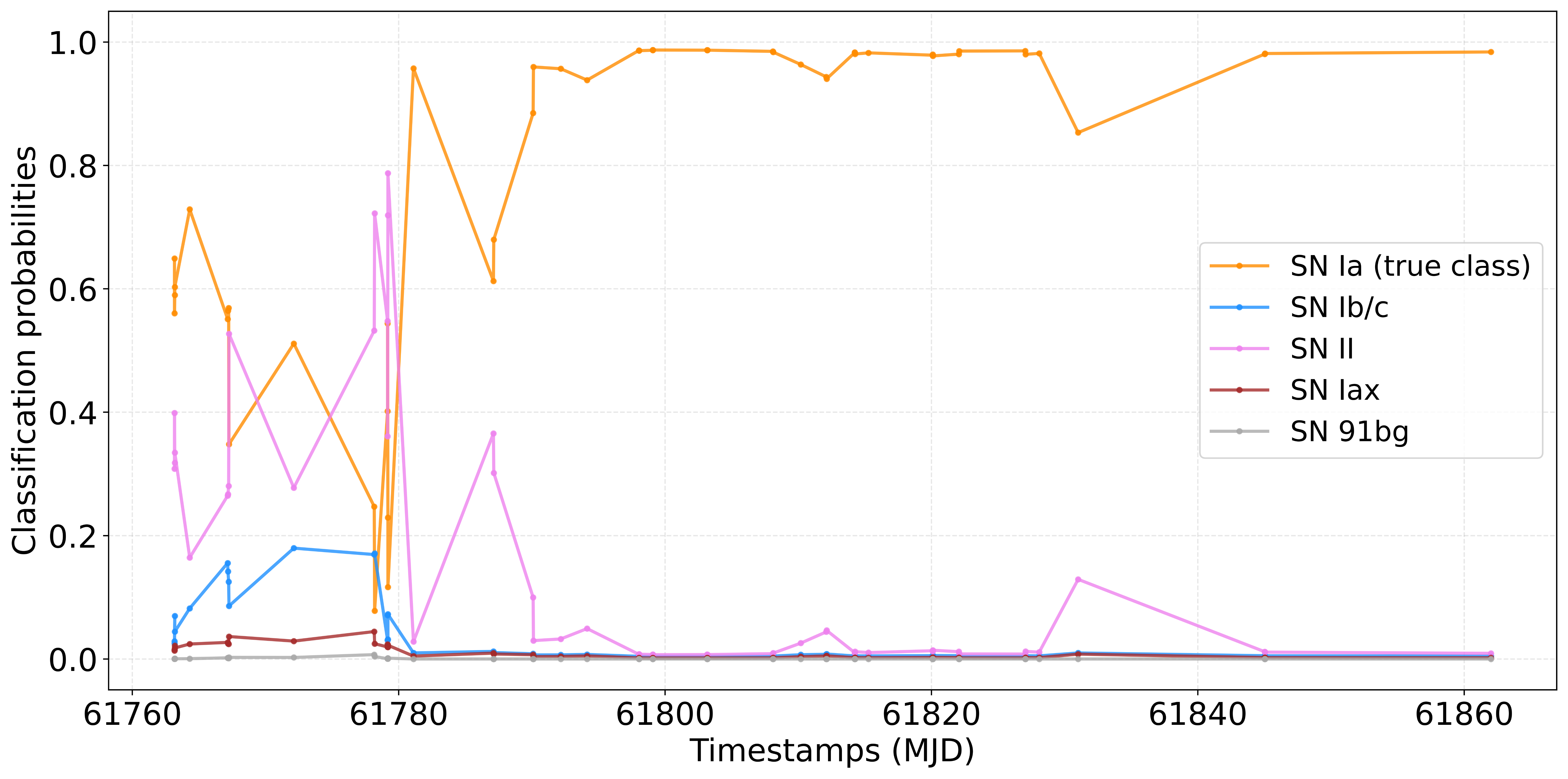} 
  \includegraphics[scale=0.45, width=1\linewidth]{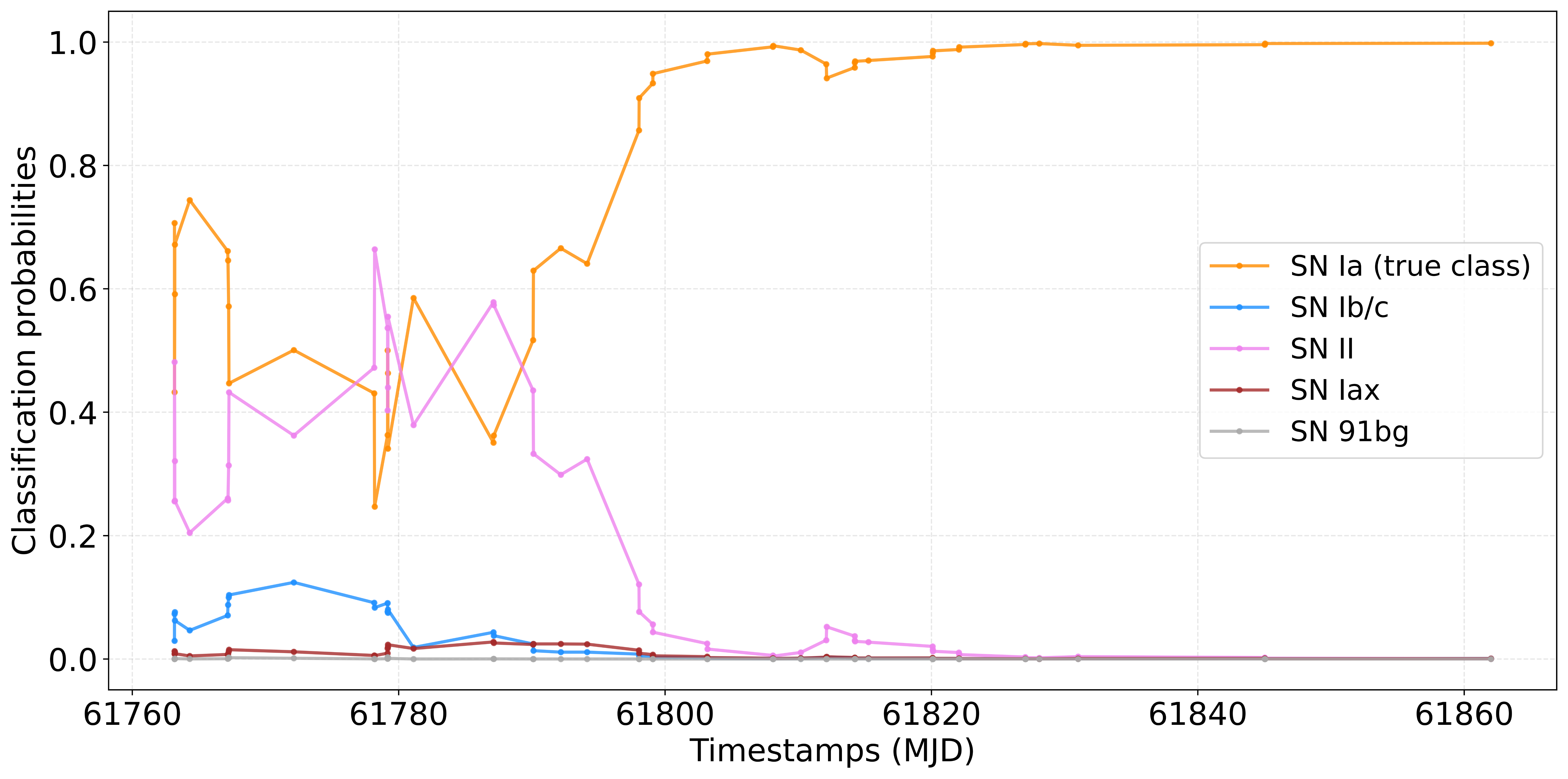}
  \caption{The comparisons of classification PMFs for an SN Ia object, object\_id=10362584, from the test dataset among baseline (upper), naive (middle), and new (bottom) models for classifier A. The new classifier acts more like a smoother or stabilizer in this case.}
  \label{fig5-1}
\end{figure}

Figure \ref{fig5-1} presents the classification PMFs produced by the three classifiers for the selected SN Ia object. 
The baseline classifier correctly classifies the object, with final classification probability closer
to one, but with significant fluctuations in probabilities at early stages. 
There is evident confusion with the SN II class. 
These fluctuations decrease, and the classifier assigns more stable and consistently high probabilities for the true class at later times. 
The naive classifier, which utilizes only the most recent classification PMFs, produces a similar temporal pattern but demonstrates reduced fluctuations and marginally improved stability compared to the baseline. 
In contrast, the new classifier generates substantially more stable classification PMFs throughout the observation sequence. At early times, the new model eliminates the drastic fluctuations exhibited by the baseline classifier. These include the swings in probability for the SN Ia class (from $<$0.2 to $>$0.8, then declining to around 0.4) and the large oscillations between SN Ia and SN II classifications. Furthermore, after MJD 61800, the classification PMFs exhibit minimal fluctuations, maintaining high stability with consistently high probabilities for the true class.

\begin{figure}
  \centering
  \includegraphics[scale=0.45, width=1\linewidth]{./figs/baseline_29_ibc.png} 
  \includegraphics[scale=0.45, width=1\linewidth]{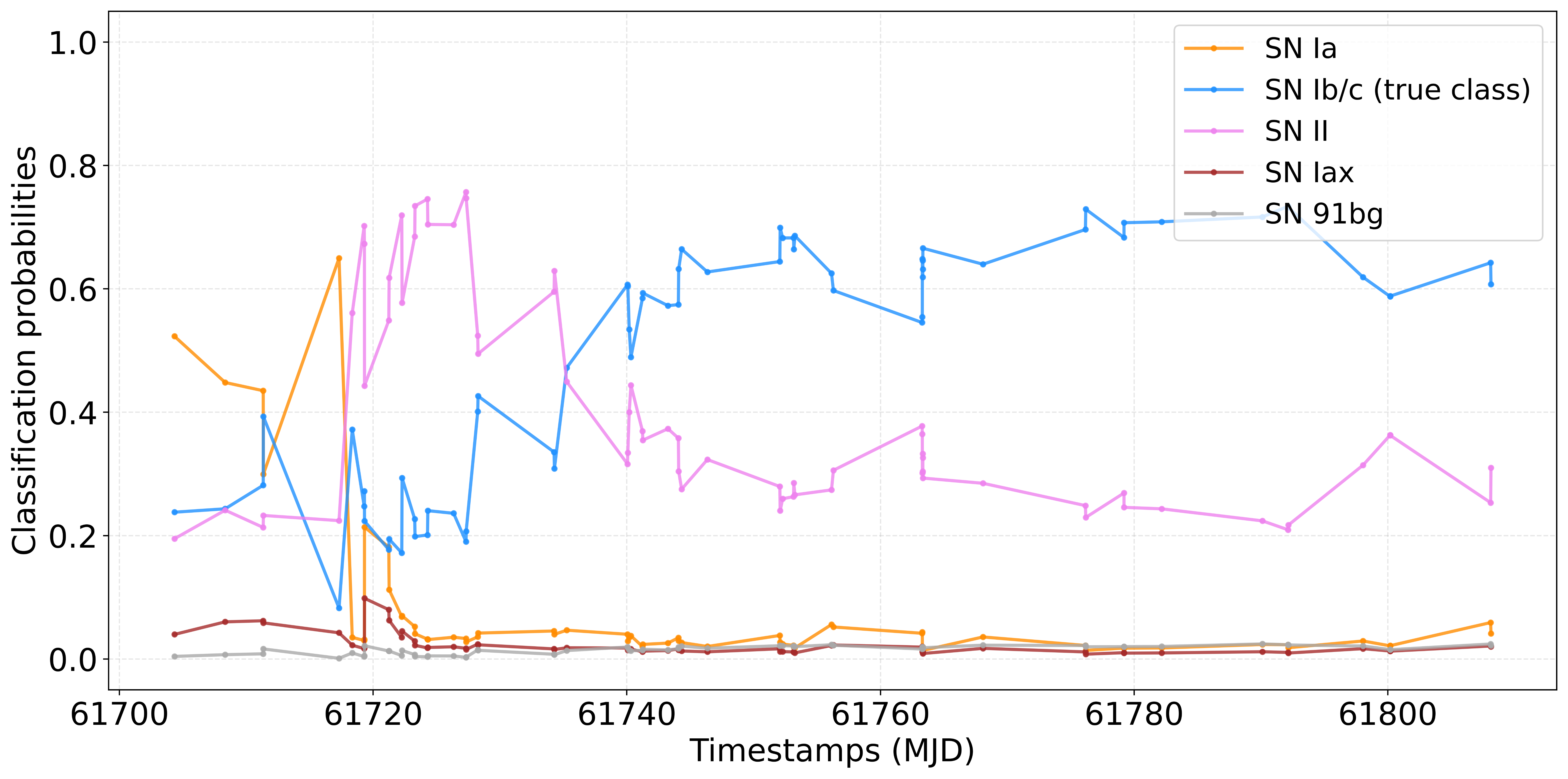} 
  \includegraphics[scale=0.45, width=1\linewidth]{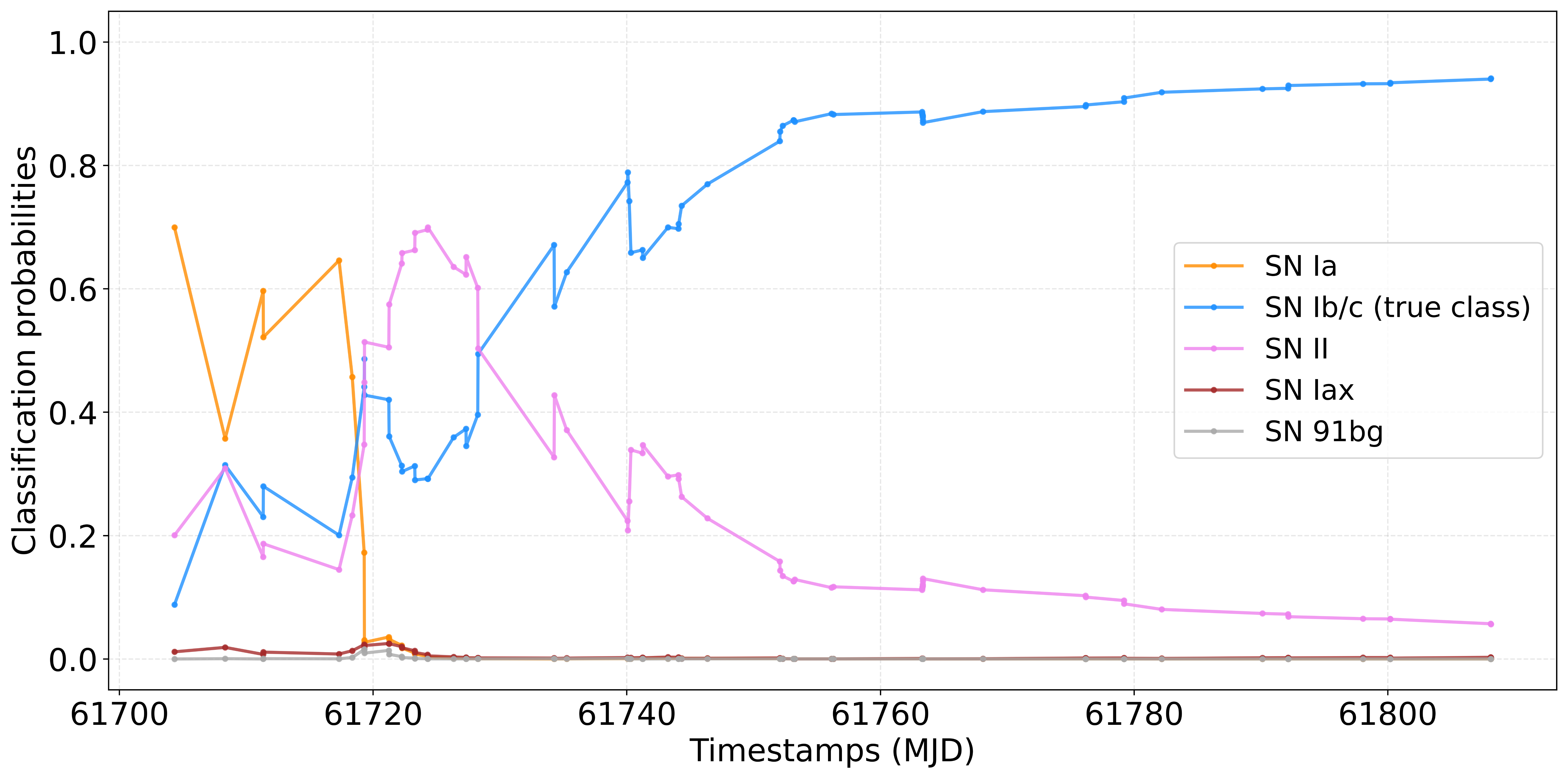}
  \caption{The comparisons of classification PMFs for an SN Ib/c test object, object\_id=1472297, from the test dataset among baseline (upper), naive (middle), and new (bottom) models for classifier A. 
  The new classifier demonstrates error and bias correction functionality.}
  \label{fig5-2}
\end{figure}

In Figure \ref{fig5-2}, we present the classification PMFs generated by the three classifiers for the selected SN Ib/c object. 
While the baseline classifier correctly classifies this object, it exhibits substantial confusion between the SN Ib/c and SN II classes throughout the classification histories, characterized by high temporal variability and instability. 
During the early times (MJD 61720–61740), the SN II probability (pink line) dominates over the true SN Ib/c class (blue line). 
Although the classification probability for the true class gradually increases, there are evident oscillations between the SN Ib/c and SN II classes. 
Notably, despite achieving correct final classification, the probabilities for both the SN Ib/c and SN II classes are near 0.5, resulting in an ambiguous and less informative classification for decision-making. 

The naive classifier addresses some of these drawbacks by producing more stable classifications with consistently higher probabilities for the true class after MJD 61740, eliminating the oscillatory behavior. 
The final classification PMFs are also more discriminative, assigning probability approximately 0.6 to the true class and 0.3 to the SN II class.
The new classifier demonstrates substantial improvement over both baseline and naive classifiers. 
After initial instability, the new classifier produces classification PMFs with consistently increasing probabilities for the true class and minimal fluctuations. 
The classification is more ideal with a stable and high probability for the true class. 
The final classification probability reaches approximately 0.9, providing significantly stronger confidence for the SN Ib/c label as compared to both the baseline and naive approaches.

The comparisons of these two selected objects provide important insights into the behavior of the proposed model. 
For objects that already achieve satisfactory classifications under the baseline classifier, such as the selected SN Ia example, the new model functions more similarly as a stabilizer, reducing temporal fluctuations while preserving classification accuracy. 
In contrast, for objects exhibiting less ideal baseline performance, characterized by inter-class confusion and high classification instability, the proposed model gives more decisive classifications by reducing or eliminating the confusion and potential misclassification in addition to the improved temporal stability.
This demonstrates the benefits of incorporating historical classifications, specifically, self-reflective error correction and the stability gains from incremental learning, as discussed in Section \ref{sec:methods}. 
Alternatively, the comparisons are only based on two selected objects for illustration. 
A more comprehensive analysis comparing a larger set of test objects would yield more decisive conclusions.

\subsection{Visualization Examples}\label{sec:visualization}

\begin{figure}
  \centering
  \includegraphics[scale=0.43, width=0.98\linewidth]{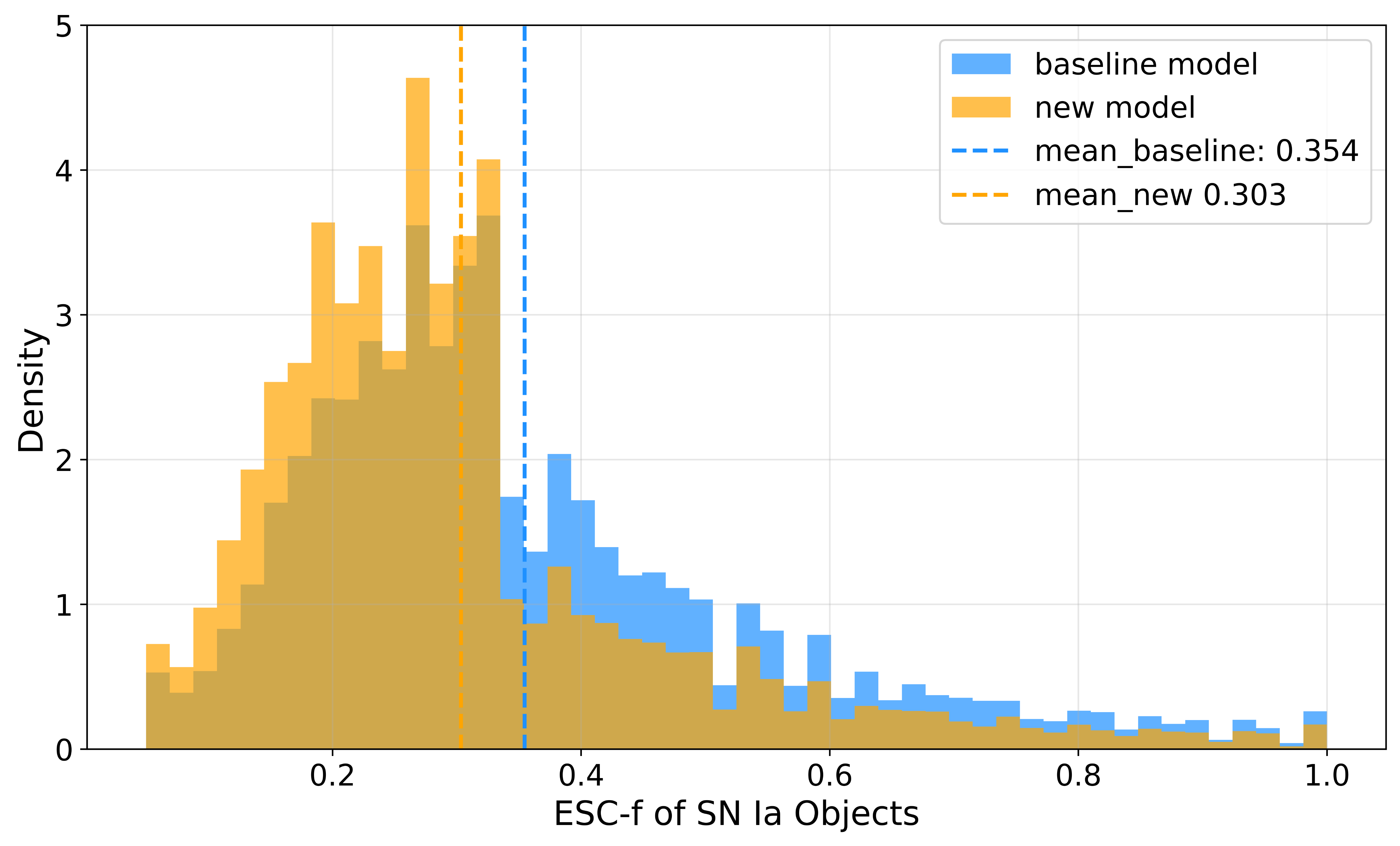} 
  \includegraphics[scale=0.43, width=0.98\linewidth]{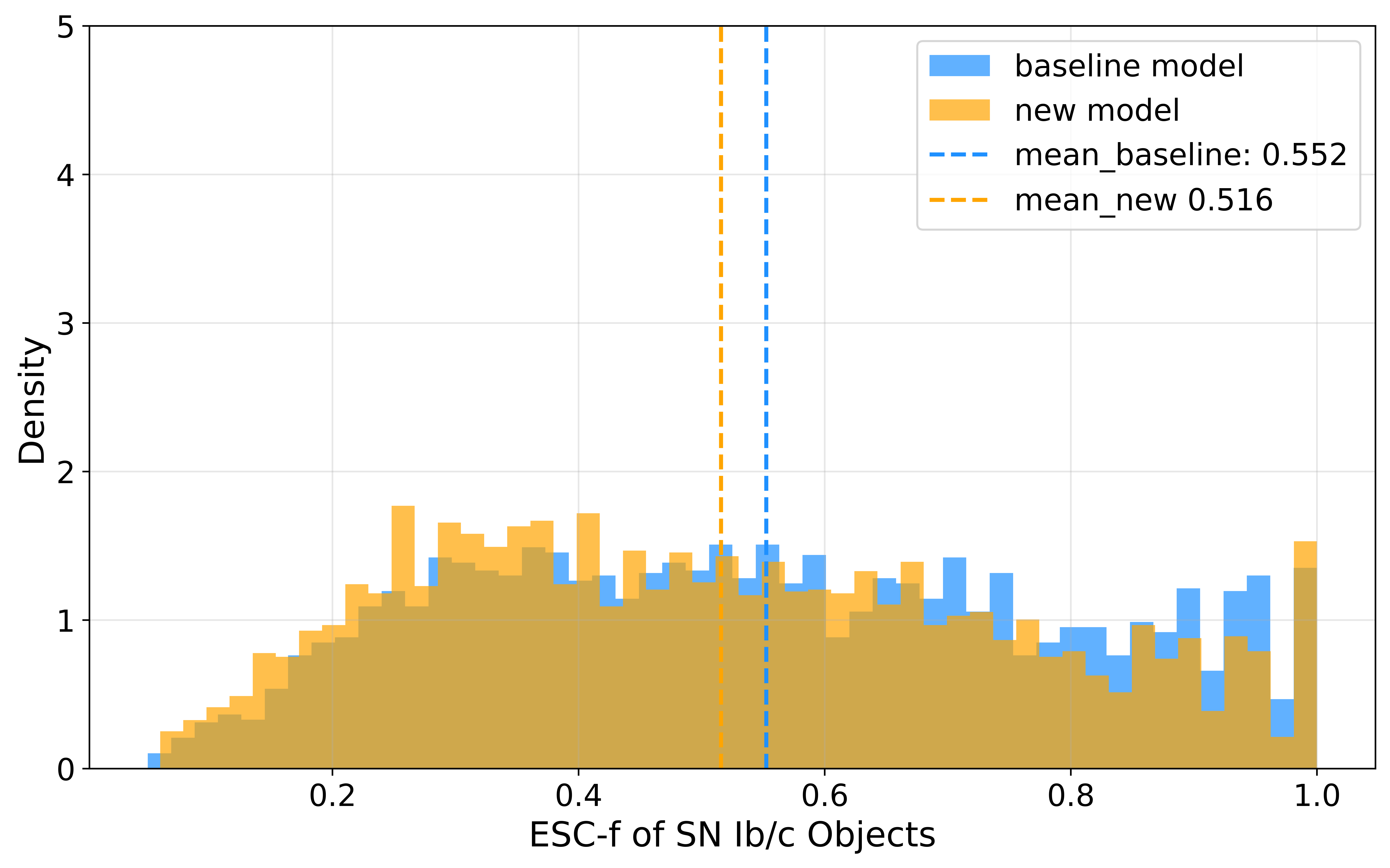} 
  \includegraphics[scale=0.43, width=0.98\linewidth]{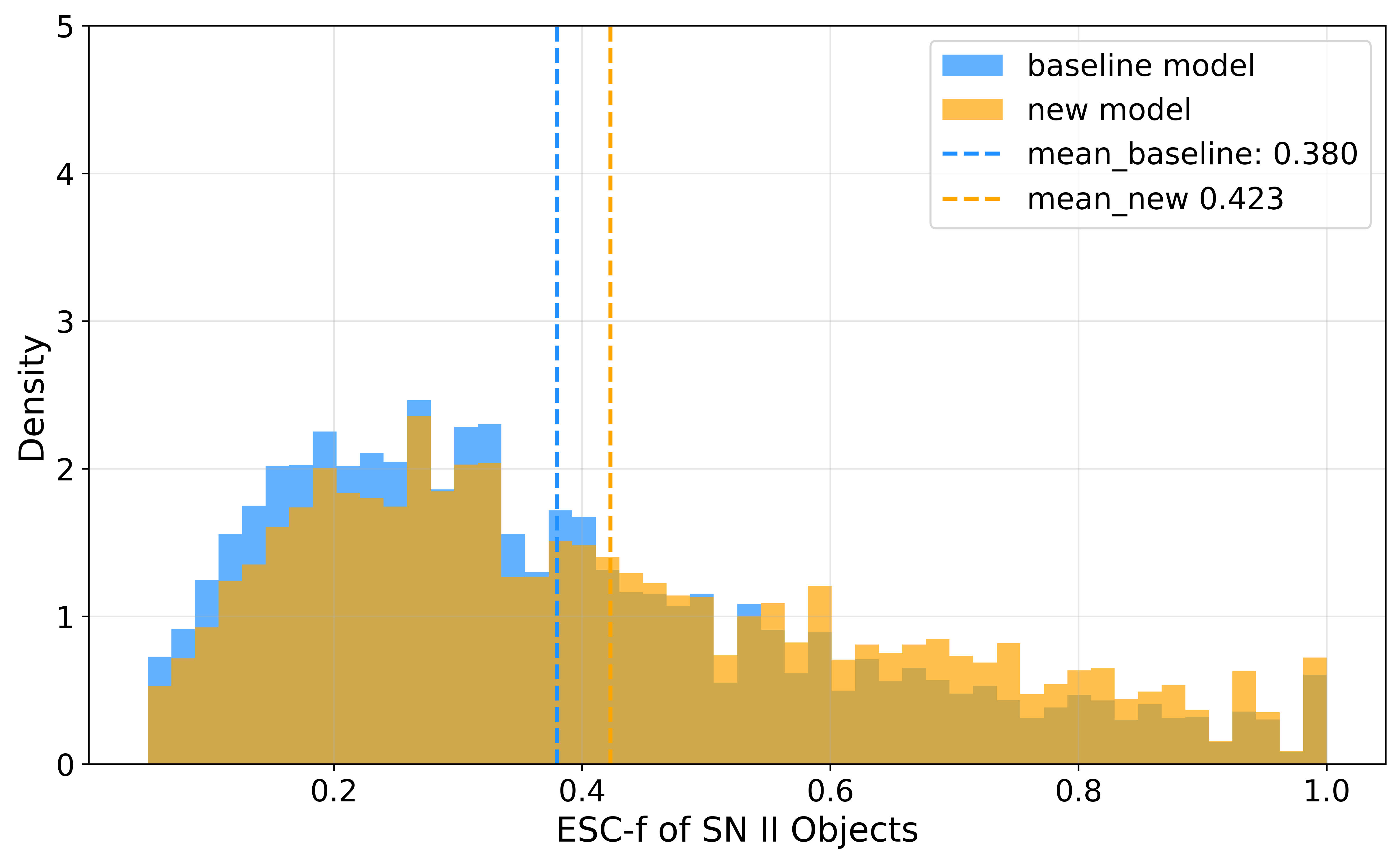}
  \caption{The comparisons of Early-Stable Classification fractions for SN Ia, SN Ib\c, and SN II objects between baseline classifier (blue) and proposed new classifier (orange) for classifier A. 
  The new classifiers achieve smaller fractions, indicating earlier convergence, for SN Ia and SN Ib/c objects, but a slightly higher fraction for SN II class.}
  \label{fig6}
\end{figure}

In this section, we visualize results using the proposed metrics for more detailed class-level comparisons. 
This serves to demonstrate how these metrics reveal detailed performance differences across individual classes.

In Figure \ref{fig6}, for classifier A, we computed the Early-Stable Classification fractions for SN Ia, SN Ib/c, and SN II, comparing the baseline and the proposed classifiers. 
We apply the same convergence conditions as in Section \ref{sec:new eval}, with $\epsilon=0.1$, $\rho=0.5$, and  $k=5$. 
For cleaner visualizations, fractions are computed only for objects that achieved convergence, rather than following the definitions of Early-Stable Classification fractions in Section \ref{sec:ecs} by assigning a value of 1 to non-converged objects. 

For SN Ia, the proposed classifier shows a leftward shift in the distribution of Early-Stable Classification fractions with a smaller mean fraction and reduced density in the right tail compared with the baseline model. 
This indicates that the proposed classifier achieves earlier stable classifications. 
The proposed classifier has improved performance (lower fractions) for SN Ib/c objects, while showing a marginally higher fraction for SN II objects. 
This slightly worse performance in SN II objects is likely due to the proposed classifier's lower recall compared to the baseline for SN II objects. 

\begin{figure}
  \centering
  \includegraphics[scale=0.43, width=0.98\linewidth]{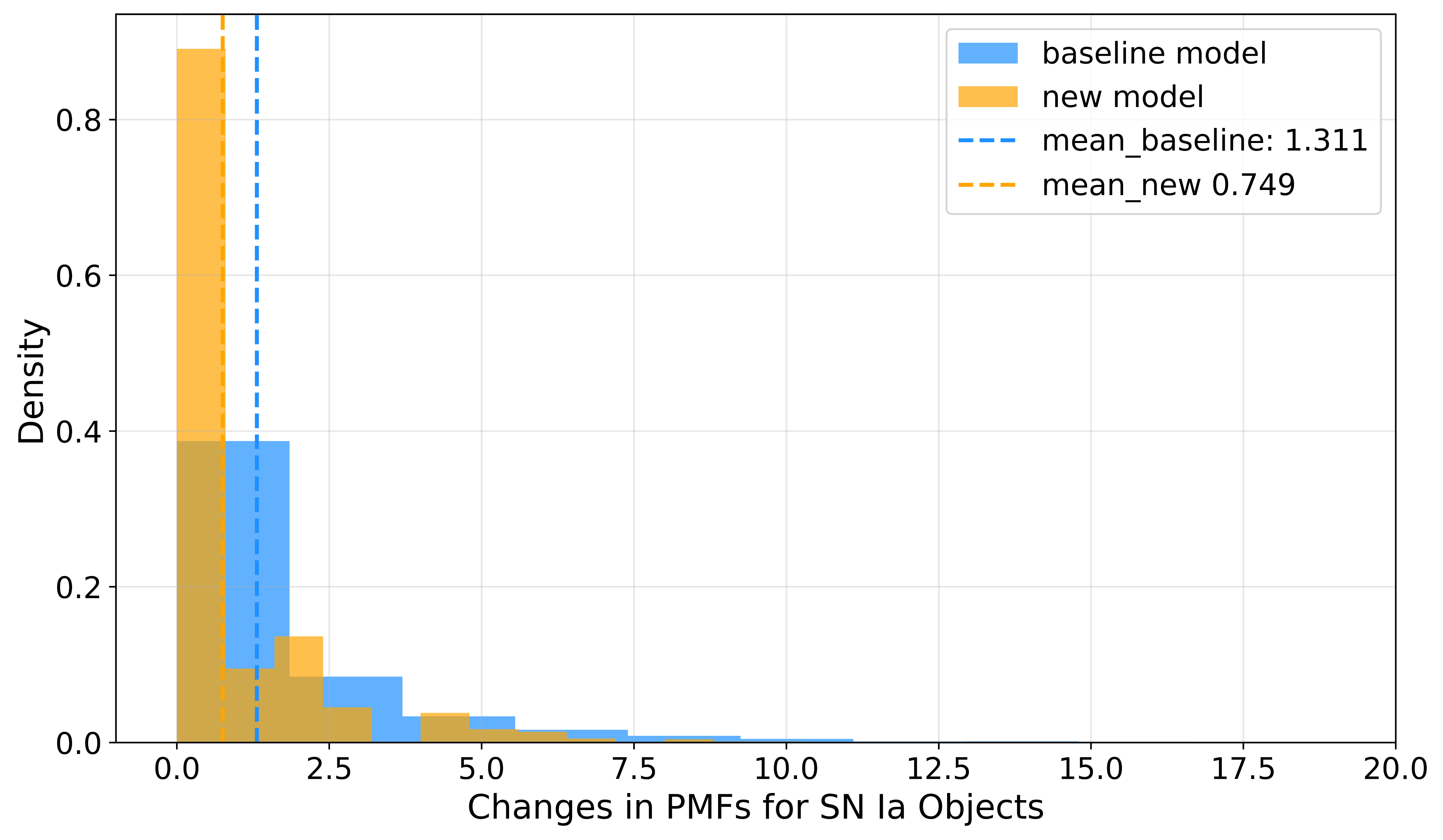} 
  \includegraphics[scale=0.43, width=0.98\linewidth]{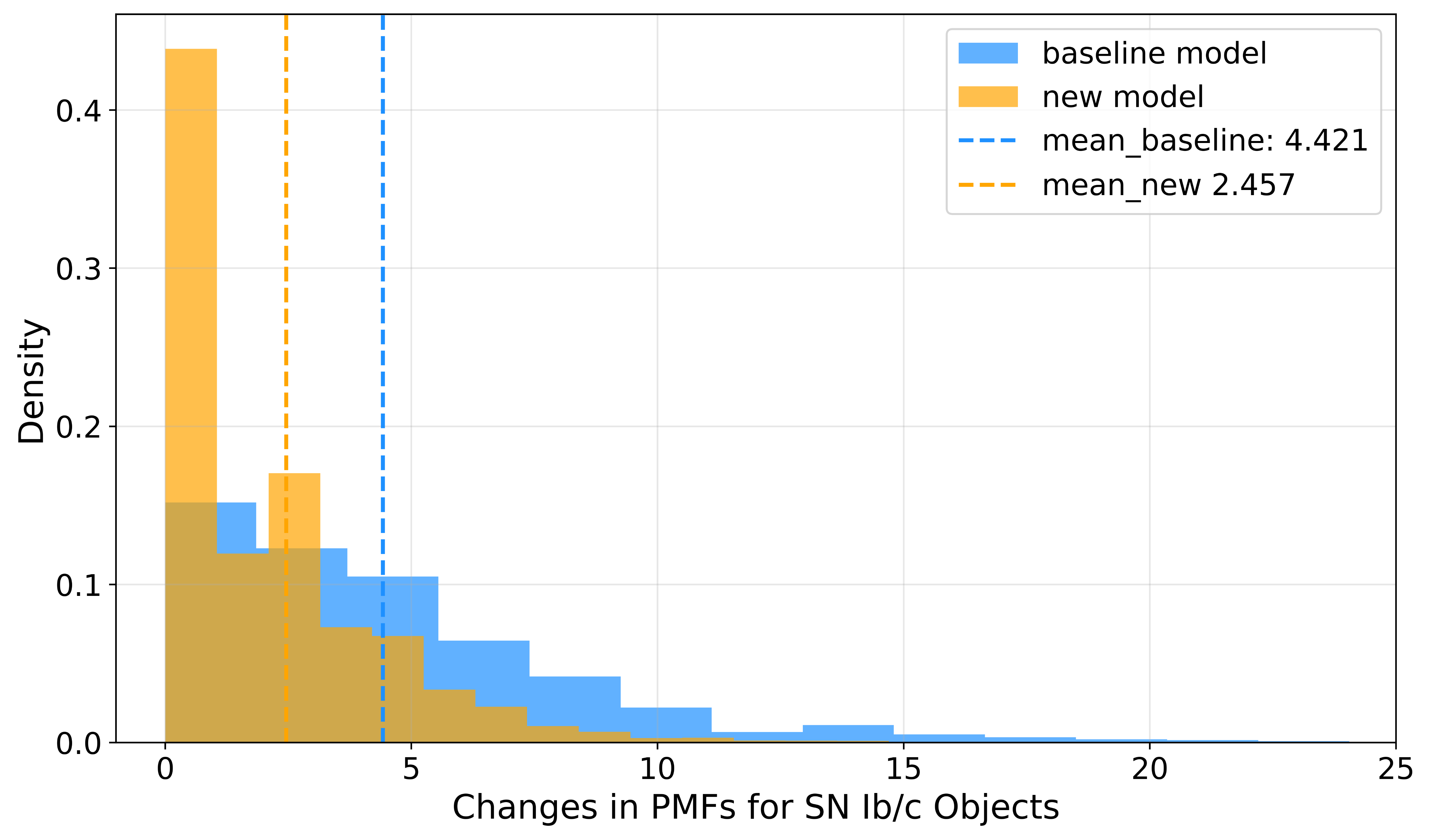} 
  \includegraphics[scale=0.43, width=0.98\linewidth]{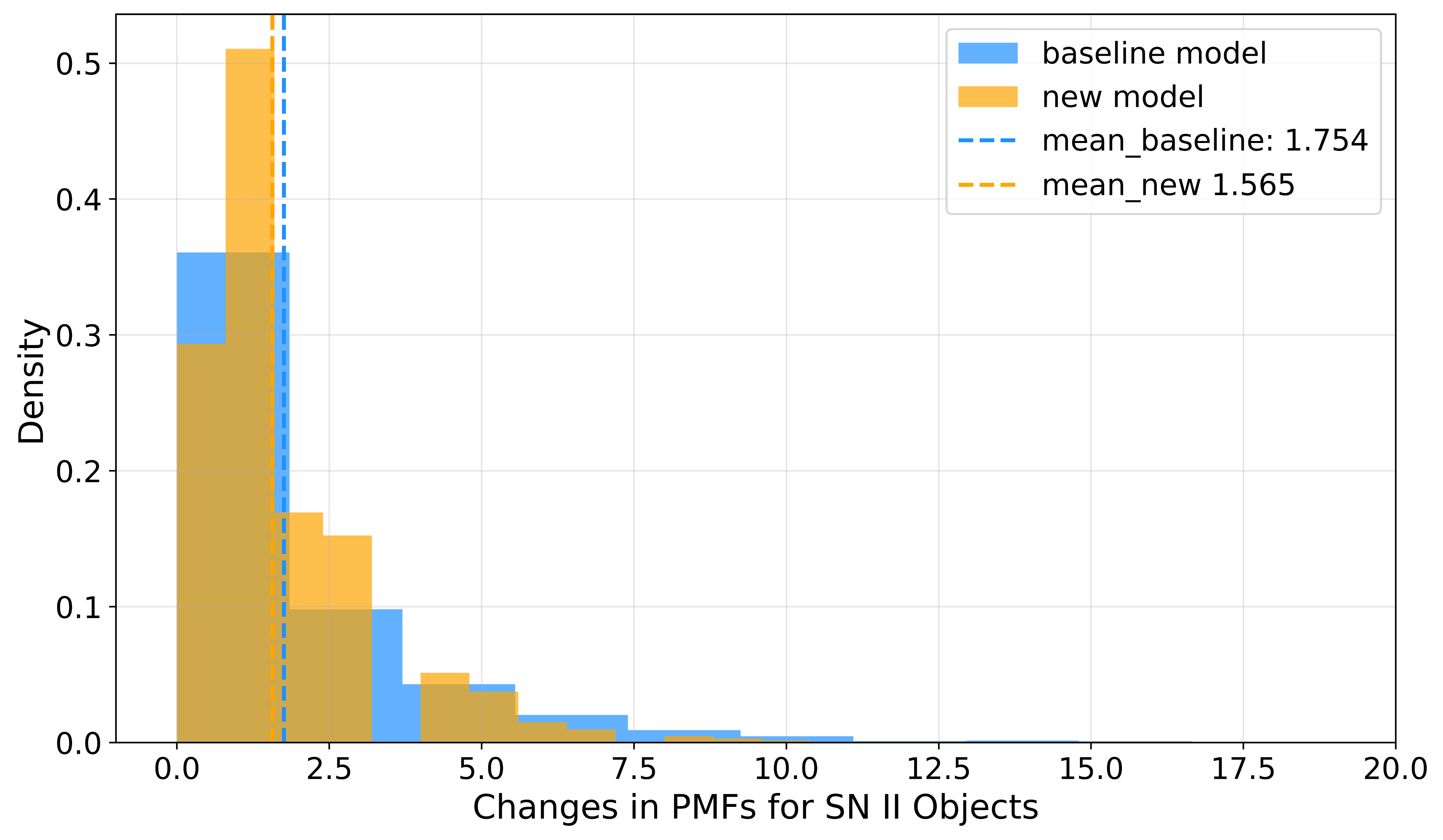}
  \caption{The comparisons of the total number of changes in classification labels for SN Ia, SN Ib/c, and SN II objects between baseline classifier (blue) and proposed new classifier (orange) for classifier A. 
  The new classifiers have a smaller number of changes for all three types of objects, indicating more stable classifications with less frequent changes in classification labels.}
  \label{fig7}
\end{figure}

For an additional intuitive and quick characterization of classification stability, we propose using the number of times the classification label (class with the highest classification probability) changes throughout the classification histories. 
In Figure \ref{fig7}, we present the distributions of average changes per class for SN Ia, SN Ib/c, and SN II, between the baseline and proposed for classifier A. 
The proposed classifier achieves a lower average number of label changes across all three classes. 
More importantly, it substantially reduces the occurrence of objects with an exceptionally high number of label changes, as suggested by the reduced right skewness in the distributions relative to the baseline classifier.

As defined in Section \ref{sec:ecs}, the Early-Stable Classification (ESC) fractions depend on user-specified parameters $(\epsilon, \rho, k)$. 
To provide a more comprehensive comparison, for classifier A, we compute the mean ESC fractions for SN Ia objects across a grid of $(\epsilon, \rho)$ while fixing $k=5$ for both the baseline and proposed models. 
We selected 10 equally spaced values for $\epsilon$ and $\rho$ between 0 and 1.

\begin{figure}
  \centering
  \includegraphics[scale=0.98, width=0.98\linewidth]{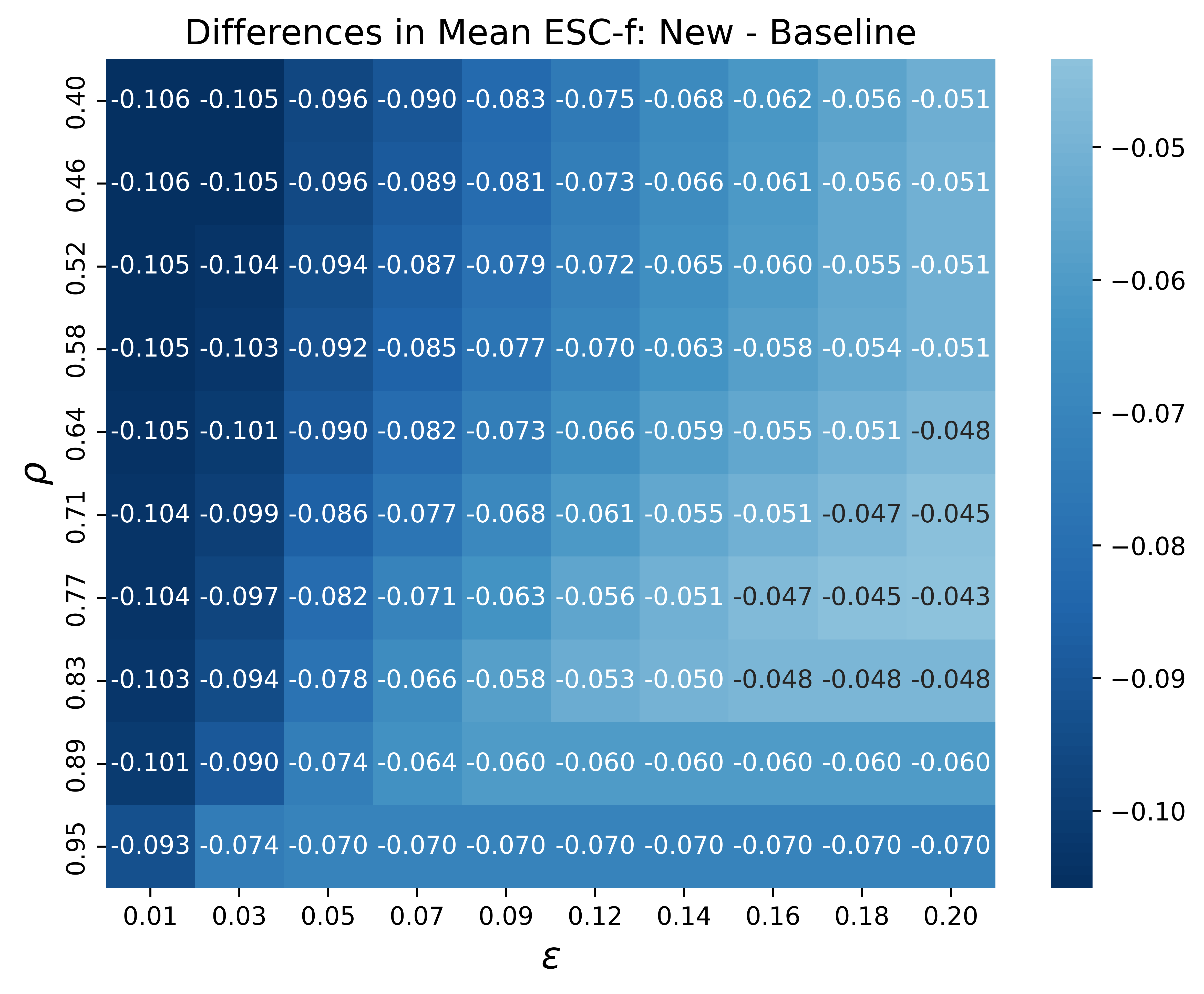} 
  \caption{Heatmap of differences in mean Early-Stable Classification fractions between baseline and new models of SN Ia objects for classifier A across a grid of $(\epsilon, \rho)$ fixing $k=5$. 
  The new classifier achieved smaller fractions across all combinations of $(\epsilon, \rho)$ and has greater improvements with more stringent conditions with smaller $\epsilon$ and larger $\rho$ (towards the bottom-left corner).}
  \label{fig8}
\end{figure}

Figure \ref{fig8} presents a heatmap where each cell represents the difference between the proposed and baseline classifiers' ESC fractions, with negative values indicating that the proposed classifier achieves better performance with earlier stable classifications. 
For SN Ia objects, the proposed classifier outperforms the baseline across all selected combinations of $(\epsilon, \rho)$. 
Notably, we have greater performance gains at smaller $\epsilon$ (stricter convergence criteria) and higher $\rho$ (higher accuracy thresholds). 
In other words, for a fixed accuracy threshold $\rho$, the proposed classifier demonstrates increasingly better early-stable performance as we impose a stricter stability requirement with a smaller $\epsilon$.

\section{Conclusions and Future Directions}\label{sec:discussion}

In this work we propose and investigate a new framework to enhance probabilistic light curve classifications by incorporating classification histories into the modeling process, motivated by the light curve classification needs of LSST. 
To demonstrate the improvements, we introduce a classifier that incorporates a Long Short-Term Memory and an additive attention mechanism to enhance existing light curve classifiers by supplementing the raw flux observations with their classification histories. 
We also define new model evaluation metrics and visualizations that better assess a classifier's early classification performance and classification stability, which are largely overlooked by existing evaluation methods. 

Using synthetic data and participating classifiers from the Extended LSST Astronomical Time-series Classification Challenge (ELAsTiCC), we demonstrate that incorporating historical classifications improves overall accuracy and achieves better precision-recall balance compared to baseline classifiers from ELAsTiCC. 
This modeling strategy also yields better early-stable classification performance. 
This demonstrates that classification histories can contribute valuable information to light curve classification and can be used to systematically evaluate light curve classifiers' early classification performance and classification stability.

We acknowledge some limitations and propose corresponding future directions. 
First, our analysis primarily focuses on five common supernovae types for demonstration purposes; 
future work should expand the scope to include additional transient or variable stars. 
Second, deeper investigation into the mechanism as to how incorporating historical classifications improves the light curve classifiers can contribute to the validity of our modeling strategy. 
Future work should explore theoretical arguments arising from stacking and ensemble learning, as well as a more detailed model evaluation by examining the misclassified objects and error correlation. 
Third, while we adopted the combination of LSTM and additive attention mechanisms, this does not represent the state-of-the-art for handling irregular time series when compared to modern transformer-based architectures. 
Given that our primary objective is to demonstrate the value of incorporating historical classifications, more sophisticated approaches for irregular time series data could be explored to further enhance classification performance. 
A particularly promising direction could adapt these metrics as loss functions for classifier training, which could lead to classifiers directly optimized for early-stable classification. 
There is a need for additional research focused on model development and evaluation techniques centered on early-stable classification. 

The experiments presented here provide a preliminary application of the proposed Early-Stable Classification Metrics metrics; immediate follow-up work could apply them to different classifiers beyond ELAsTiCC as well as in experimental settings for testing and refining them as needed. For instance, a possible limitation in our current setting concerns the origin of the classification instability due to uncertainties from light curve observations and from model's classification with partially observed data. The temporal variability in the classification PMFs (e.g. Figures \ref{fig5-1} and \ref{fig5-2}) reflects a combination of the light curve becoming more complete over time and the measurement uncertainty of the individual flux observations that enter each update. Our analysis quantifies the overall instability through the Wasserstein distance between successive PMFs but cannot attribute it to these two contributions separately. A natural way to probe this would be to draw multiple realizations of each light curve by resampling the flux at every epoch according to its reported uncertainty, propagate each realization through the classifier, and characterize the resulting evolution of PMFs. This would require re-running the classifiers on resampled inputs. However, our work relies on pre-computed PMFs without access to the base classifiers for fitting resampled light curves and is therefore deferred for future work that has better access to the baseline classification models.

Finally, since LSST is just now yielding its first observations, our work relies on high-fidelity synthetic data. 
The framework should also be validated on existing real observations from surveys such as the Zwicky Transient Facility (ZTF) Source Classification Project \citep{healy2024ztf}, and reevaluated once sufficient LSST data and classifications become available.

\begin{acknowledgments}
A.I.M., K.M., and C.M.S.~were supported by Schmidt Sciences.  
This work used Bridges-2 at Pittsburgh Supercomputing Center through allocation [PHY250104, PHY210095] from the Advanced Cyberinfrastructure Coordination Ecosystem: Services \& Support (ACCESS) program \citep{boerner2023access}, which is supported by National Science Foundation grants \#2138259, \#2138286, \#2138307, \#2137603, and \#2138296. G.C.V. gratefully acknowledges financial support from ANID: FONDECYT Regular grant 1231877 and MILENIO NCN2024112. The authors thank Rob Knop for enabling the public access of the data.

\end{acknowledgments}

\software{The Jupyter Notebook files and code for the proposed models in Section \ref{sec:methods} and model training are available at \url{https://github.com/grantzzhou/LSST_beyond_the_final_label} and the Zenodo repository \url{https://zenodo.org/records/20261776} \citep{zhou_2026_20261776}. The light curves data used for model training and evaluation is available at \url{https://zenodo.org/records/20261776}. All code is implemented in Python, with PyTorch used for building and training the model.}

\begin{contribution}
Z.Z.~led the analysis, writing, interpretation, and modeling of the paper. A.I.M.~and C.M.S.~propose and led the project, oversaw the project progress, and contributed to the analysis, interpretation, and writing. K.M.~led the ELAsTiCC data pre-processing and contributed to reviewing of the paper. G.C.V.~and C.H.~are the core developers of the alert brokers used in the project and contributed to reviewing of the paper. 
    
\end{contribution}

\appendix
\restartappendixnumbering
\section{Supplemental Model Evaluations}\label{sec:appendix}

\begin{figure*}
  \centering
  \includegraphics[scale=0.45]{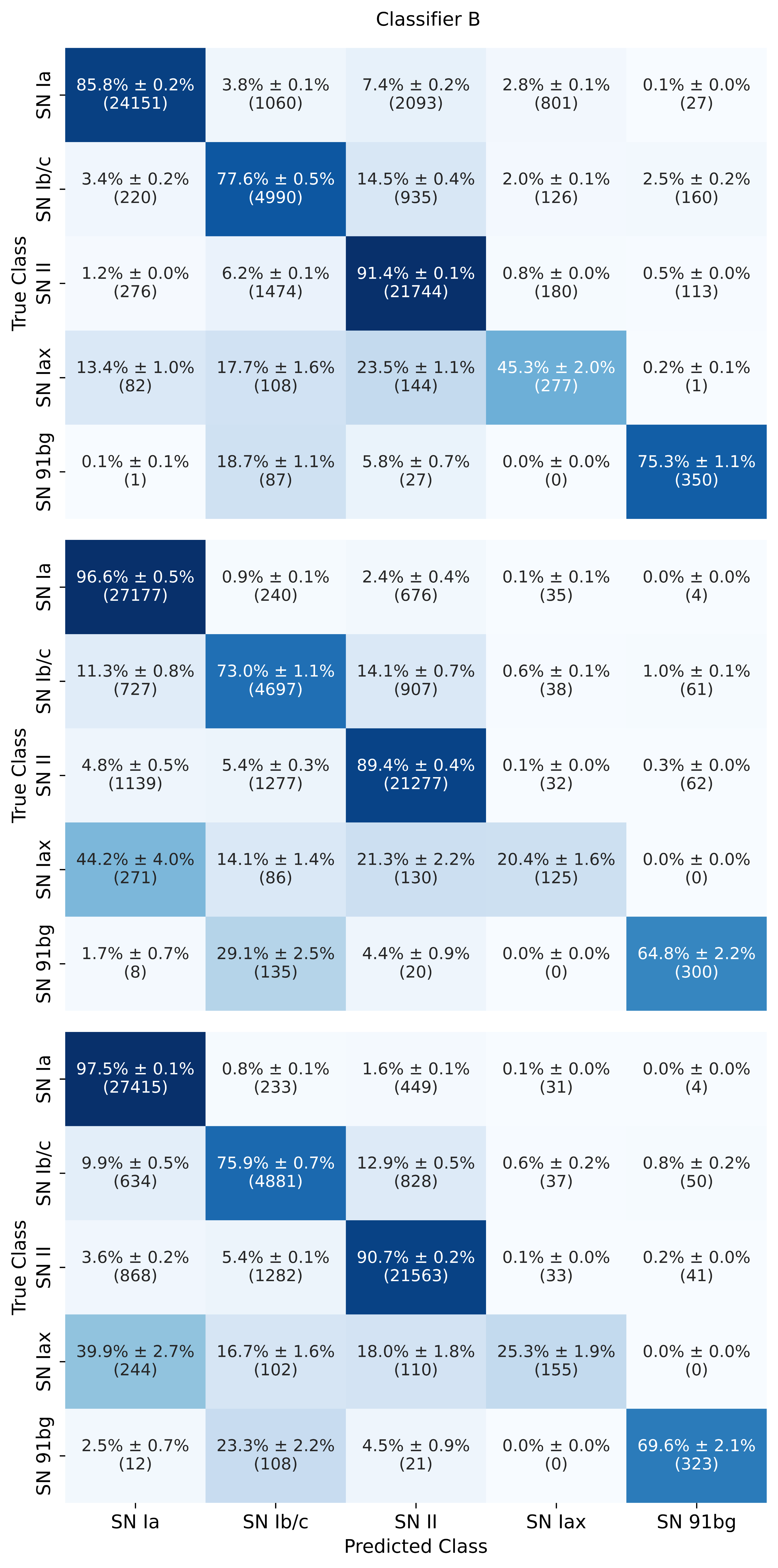} 
  \caption{The comparisons between the baseline (upper row), naive (middle), and new (bottom row) classifiers for classifier B. 
  The confusion matrix is normalized per row and annotated with average absolute counts. 
  The new models show improvements in overall accuracy and more balanced precision-recall.}
  \label{fig-sup}
\end{figure*}

In this section, we present the average confusion matrices for classifier B (Figure \ref{fig-sup}). As mentioned in Section \ref{sec:eval classic}, we omit discussion of the confusion matrices for classifier B as this classifier originates from the same alert broker as classifier A and exhibits similar behavior due to space constraints. 

We also provide the detailed proportions of convergence and Early-Stable Classification fractions for one of the ten train-test splits (random seed $=0$) for the three selected classifiers and across the three model types (Tables \ref{sup-table-a}, \ref{sup-table-b}, \ref{sup-table-c}). 
Note that the variations across the splits are relatively small and will not affect any conclusions, so we only select one of the splits instead of presenting results from all splits.

\newpage
\begin{table*}
  \caption{Detailed Evaluation for Classifier A}
  \centering
  \setlength{\tabcolsep}{4pt}
  \renewcommand{\arraystretch}{0.9} 
  \begin{tabular}{lc|ccccc}
    \toprule
    & & \multicolumn{5}{c}{\textbf{Object Class}} \\
    \cmidrule(lr){3-7}
    \textbf{Metric} & \textbf{Model} & \textbf{SN Ia} & \textbf{SN Ib/c} & \textbf{SN II} & \textbf{SN Iax} & \textbf{SN 91bg} \\
    \midrule
    \multirow{3}{*}{\shortstack[l]{Convergence \\ Proportion}}
    & Baseline & 0.918 & 0.460 & 0.865 & 0.0294 & 0.448\\
    & Naive & 0.925 & 0.576& 0.0216 & 0.000778 & 0.000178\\
    & New & 0.963 & 0.640 & 0.856 & 0.127 & 0.534\\
    \midrule
    \multirow{3}{*}{\shortstack[l]{Early-Stable\\ Classification Fraction}}
    & Baseline & 0.404 & 0.789 & 0.457 & 0.987 & 0.793\\
    & Naive & 0.417 & 0.711 & 0.988 & 1.000 & 1.000 \\
    & New & 0.327 & 0.681 & 0.499 & 0.959 & 0.759\\
    \midrule
    \multirow{3}{*}{\shortstack[l]{Early-Stable\\ Classification Score}} 
    & Baseline & 0.241 & 1.078 & 0.321 & 4.048 & 1.843 \\
    & Naive & 0.261 & 0.907 & 3.429 & 4.520 & 10.099\\
    & New & 0.241 & 0.874 & 0.464 & 2.924 & 1.945\\
    \midrule
    \multirow{3}{*}{\shortstack[l]{Average\\Changes}} 
    & Baseline & 1.311 & 4.421 & 1.754 & 4.150 & 3.935\\
    & Naive & 1.135 & 3.957 & 1.135 & 3.957 & 1.135 \\
    & New & 0.749 & 2.457 & 1.565 & 2.529 & 2.774\\
    \bottomrule
  \end{tabular}
  \label{sup-table-a}
\end{table*}

\begin{table*}
  \caption{Detailed Evaluation for Classifier B}
  \centering
  \setlength{\tabcolsep}{4pt}
  \renewcommand{\arraystretch}{0.8} 
  \begin{tabular}{lc|ccccc}
    \toprule
    & & \multicolumn{5}{c}{\textbf{Object Class}} \\
    \cmidrule(lr){3-7}
    \textbf{Metric} & \textbf{Model} & \textbf{SN Ia} & \textbf{SN Ib/c} & \textbf{SN II} & \textbf{SN Iax} & \textbf{SN 91bg} \\
    \midrule
    \multirow{3}{*}{\shortstack[l]{Convergence \\ Proportion}}
    & Baseline & 0.730 & 0.638 & 0.834 & 0.245 & 0.703\\
    & Naive & 0.923 & 0.611 & 0.04 & 0.00156 & 0.00174\\
    & New & 0.963 & 0.675 & 0.851 & 0.132 & 0.509\\
    \midrule
    \multirow{3}{*}{\shortstack[l]{Early-Stable\\ Classification Fraction}}
    & Baseline & 0.630 & 0.671 & 0.489 & 0.887 & 0.570\\
    & Naive & 0.413 & 0.678 & 0.975 & 0.999 & 0.999\\
    & New & 0.369 & 0.655 & 0.509 & 0.957 & 0.765\\
    \midrule
    \multirow{3}{*}{\shortstack[l]{Early-Stable\\ Classification Score}} 
    & Baseline & 0.609 & 0.685 & 0.393 & 1.628 & 0.907\\
    & Naive & 0.309 & 0.867 & 3.712 & 5.087 & 11.110\\
    & New & 0.254 & 0.816 & 0.492 & 2.911 & 1.849 \\
    \midrule
    \multirow{3}{*}{\shortstack[l]{Average\\Changes}} 
    & Baseline & 3.245 & 3.994 & 2.075 & 5.451 & 2.394\\
    & Naive & 1.608& 4.0253 & 1.608 & 4.0253 & 1.608\\
    & New & 0.923 & 2.342 & 1.551 & 2.541 & 2.302\\
    \bottomrule
  \end{tabular}
  \label{sup-table-b}
\end{table*}

\begin{table*}
  \caption{Detailed Evaluation for Classifier C}
  \centering
  \setlength{\tabcolsep}{4pt}
  \renewcommand{\arraystretch}{0.8} 
  \begin{tabular}{lc|ccccc}
    \toprule
    & & \multicolumn{5}{c}{\textbf{Object Class}} \\
    \cmidrule(lr){3-7}
    \textbf{Metric} & \textbf{Model} & \textbf{SN Ia} & \textbf{SN Ib/c} & \textbf{SN II} & \textbf{SN Iax} & \textbf{SN 91bg} \\
    \midrule
    \multirow{3}{*}{\shortstack[l]{Convergence \\ Proportion}}
    & Baseline & 0.528 & 0.513 & 0.539 & 0.460 & 0.660\\
    & Naive & 0.780 & 0.368 & 0.104 & 0.000 & 0.00457 \\
    & New & 0.768 & 0.491 & 0.823 & 0.0336 & 0.203\\
    \midrule
    \multirow{3}{*}{\shortstack[l]{Early-Stable\\ Classification Fraction}}
    & Baseline & 0.792 & 0.777 & 0.763 & 0.763 & 0.673\\
    & Naive & 0.627 & 0.850 & 0.941 & 1.000 & 0.997\\
    & New & 0.713 & 0.826 & 0.520 & 0.994 & 0.957\\
    \midrule
    \multirow{3}{*}{\shortstack[l]{Early-Stable\\ Classification Score}} 
    & Baseline & 1.026 & 0.864 & 0.767 & 1.041 & 0.794\\
    & Naive & 0.505 & 1.224 & 2.192 & 5.527 & 10.400\\
    & New & 0.736 & 1.211 & 0.404 & 3.002 & 2.638 \\
    \midrule
    \multirow{3}{*}{\shortstack[l]{Average\\Changes}} 
    & Baseline & 3.300 & 3.596 & 3.138 & 3.437 & 2.499\\
    & Naive & 2.149 & 3.861 & 2.149 & 3.861 & 2.149 \\
    & New & 2.438 & 3.179 & 1.516 & 3.109 & 4.080\\
    \bottomrule
  \end{tabular}
  \label{sup-table-c}
\end{table*}

\newpage
\clearpage  
\bibliographystyle{aasjournal}
\bibliography{ref.bib} 

\end{CJK*}
\end{document}